\documentclass{article}
\usepackage[utf8]{inputenc}
\usepackage{amsthm,amsmath,amssymb}
\usepackage{graphicx}
\usepackage{authblk,setspace}
\usepackage[shortlabels]{enumitem}
\usepackage[a4paper, total={6in, 8in}]{geometry}
\usepackage{lscape}
\usepackage{booktabs}
\usepackage{moreverb,url,setspace}

\title{Evaluating the impact of outcome delay on the efficiency of two-arm group-sequential trials}
\author[1]{Aritra Mukherjee}
\author[2]{Michael J. Grayling}
\author[1]{James M. S. Wason}

\affil[1]{Population Health Sciences Institute, Newcastle University}
\affil[2]{Janssen R\&D}

\date{}

\begin{document}

\onehalfspacing
\maketitle


\begin{abstract}
\noindent \textbf{Background}: Adaptive designs are a broad class of designs that allow modifications to be made to a trial as participant data is accrued.
Adaptive designs can offer improved efficiency and flexibility.
However, a delay in observing the primary outcome variable can potentially harm the efficiency added by adaptive designs.
Principally, in the presence of such delay, we may have to make a choice as to whether to (a) pause recruitment until requisite data is accrued for the interim analysis, leading to a longer study duration; or (b) continue to recruit participants, which may result in a large number of participants who do not benefit from the interim analysis and damage to the study's efficiency.
Little work has been conducted to ascertain the size of outcome delay that results in the realised efficiency gains of adaptive designs becoming negligible compared to classical fixed-sample alternatives.
We perform such work here for two-arm group-sequential designs (GSDs) with different numbers of interim analyses.

\noindent \textbf{Methods:} We measure the impact of outcome delay by developing formulae for the number of `pipeline' participants (sometimes called `overruns') in two-arm GSDs with normal data, assuming different recruitment models.
Typically, the efficiency of a GSD is measured in terms of the expected sample size (ESS), with GSDs generally reducing the ESS compared to a design without interim analyses.
Our formulae enable us to measure the efficiency gain from a GSD in terms of ESS reduction that is lost due to outcome delay.
We assess whether careful choice of design (e.g., altering the spacing of the interim analyses) can help recover the advantages of GSDs in the presence of outcome delay.
We similarly analyse the efficiency of GSDs with respect to time to complete the trial. 

\noindent \textbf{Results:} On comparing the expected efficiency gains, with and without consideration of the impact of delay, it is evident GSDs can suffer considerable losses due to outcome delay.
Even a small delay can have a considerable impact on the trial's efficiency in terms of an increased EL.
On the other hand, even in the presence of substantial delay, a GSD will have a smaller expected time to trial completion in comparison to a single-stage trial.
With increase in the number of stages the efficiency loss increases. 
The timing of the IAs can further considerably impact the efficiency of a GSD for delayed outcomes; in particular, conducting IAs too early in the trial can harm the design's efficiency.

\noindent \textbf{Conclusions:} A delay in observing the treatment outcome is harmful to the efficiency of a GSD.
Even a small delay in presence of an increasing recruitment pattern can incur heavy losses to the design.
Furthermore, with unequally spaced interims, pushing analyses towards the latter end of the trial can save some of the lost efficiency for the design in the presence of delay.


\end{abstract}

\maketitle
\onehalfspacing
\section{Introduction}

Group-sequential designs (GSDs) are commonly used in practice for two-arm randomised controlled trials (RCTs), particularly in the later phases of drug development \cite{hatfield2016,bothwell2018}.
A GSD introduces interim analyses (IAs) that allow early termination for efficacy or futility based on the accumulating data \cite{jennison2000,grayling2018,kelly2005,wason2015}.
They have the potential to improve efficiency (e.g., in terms of the study's expected time to completion or required sample size) considerably compared to a classical design with a single analysis.
Further, as the number of stages increases, a greater efficiency gain is expected due to increased reduction in the expected sample size (ESS) \cite{jennison2000,wason2012}. 

However, long-term endpoints can heavily impact the potential advantages of GSD.
A delay in observing a treatment outcome will often either (a) increase the time to complete the trial if recruitment is paused at each IA to await collection of all outcome data, or (b) inflate the cost of the trial by recruiting a potentially larger number of participants than needed if recruitment is continued \cite{wason2019}.
For example, consider a trial that is testing a drug against the existing standard of care with 80\% power and 5\% significance level[two-sided] for a standardised effect size of 0.4.
The primary outcome is measured after one year from starting the treatment.
Then for a three-stage GSD using O'Brien and Fleming stopping boundaries \cite{obrien1979}, the required sample size at stage 1, 2 and 3 is 66, 134 and 200 respectively, with the corresponding single stage sample size being 196 participants.
If the trial aims to complete recruitment by 2 years, then the required rate of recruitment is approximately 8 participants per month.
Assuming 8 participants are recruited per month, then at the first IA, by the time outcome data is available from the first 66 participants, the trial would have recruited an additional 96 participants if recruitment is not paused.
If the trial stops at the IA, then these 96 participants were enrolled and treated needlessly.
As the time to observe the primary outcome becomes larger (or the recruitment rate becomes faster), this would lead to an even less efficient design.

Recently, Mukherjee et al. \cite{mukherjee2022} evaluated the impact of outcome delay on two-stage single-arm trials, as are commonly employed in phase 2 oncology trials.
It was observed that Simon's two-stage design may be preferable over a corresponding single-stage alternative if the ratio of the delay period to the recruitment length is less than 0.25.  
Extension of this work to randomised two-arm GSDs (which may more typically have more than two stages) is needed, despite some previous research in this area. 
In particular, Hampson and Jennison \cite{hampson2013} discussed outcome delay within the context of two-arm GSD. 
They described how delay can impact a GSD with equally spaced IAs, when participant recruitment follows a Poisson arrival process. 
They proposed a sequential test structure incorporating the outcome delay and further provided an optimal test to minimise the ESS. 
Nonetheless, they also noted out that the benefits of lower ESS that are normally achieved by a GSD are reduced when there is a delay in outcome accrual, even when an optimal design is used. 
While Hampson and Jennison provided a comprehensive theoretical solution for delayed outcomes, but to our knowledge, it is not one that is typically utilised in practice. 
At the planning stage the impact of delay on the trial’s efficiency is often overlooked and the trial is designed under the assumption of immediate treatment outcome. 
Further work is therefore required to explore how the delay length and recruitment speed impacts on efficiency, and how this is affected by the number and spacing of interim analyses.

In this study, we aim to clearly quantify the loss in efficiency provided by a GSD for a given delay in the treatment outcome. 
We provide a framework to estimate the possible number of ‘pipeline’ participants through different realistic recruitment models. 
We also seek to examine how different numbers of IAs, and the timing of these IAs, can impact the efficiency of the design. 
In addition to considering the ESS, we study the impact of outcome delay on the expected time to trial completion when using a GSD.

\section{Methods}

\subsection{Design and notation}

We consider a two-arm GSD for testing superiority of an experimental treatment over a control.
Let $n_{0k}$ and $n_{1k}$ denote the cumulative sample size at stage $k$, $k = 1, 2, \dots, K$, for the control and treatment arms respectively.
Thus we assume the design has at most $K$ stages.
Further, let $n_k = n_{0k} + n_{1k}$.

For illustration, we assume the treatment response from participant $i = 1, 2, \dots, n_{jK}$ in arm $j = 0, 1$ is distributed as $X_{ij} \sim N\left(\mu_j, \sigma_j^2\right)$, with $\sigma_0$ and $\sigma_1$ known.
Extension to many other types of outcome (e.g., binary, count) follows naturally if test statistics follow the canonical joint distribution described by Jennison and Turnbull\cite{jennison2000}.
We suppose the trial is powered to test the hypothesis $H_0 : \mu \le 0$ against $H_1 : \mu > 0$, for $\mu = \mu_1 - \mu_0$, at significance level $\alpha$, with power $1 - \beta$ when $\mu = \tau > 0$.

At IA $k$, the test statistic used is

$$ Z_k = \frac{\frac{1}{n_{1k}} \sum_{i = 1}^{n_{1k}} X_{i1} - \frac{1}{n_{0k}} \sum_{i = 1}^{n_{0k}} X_{i0}}{\sqrt{\frac{\sigma_0^2}{n_{0k}} + \frac{\sigma_1^2}{n_{1k}}}}. $$

The GSD is assumed to use efficacy and (binding) futility stopping boundaries. 
There are many approaches available to determine these stopping boundaries, including Pocock's \cite{pocock1977}, O'Brien-Fleming's \cite{obrien1979}, and Wang-Tsiatis'\cite{wang1987} methods.
Alternatively, an $\alpha$-spending approach may be adopted, where the boundaries at stage $k$ are dependent on the proportion, $\rho_k$, of the maximal Fisher's information that is available at IA $k$.
If we denote the efficacy and futility boundaries used at analysis $k$, determined by a given method, by  $e_k$ and $f_k$, then the following stopping rules are used
\begin{itemize}
    \item stop at IA $k$ for efficacy if $Z_k > e_k$;
    \item stop at IA $k$ for futility if $Z_k \le f_k$;
    \item continue to IA $k + 1$ if $f_k < Z_k \le e_k$.
\end{itemize}

Next, define

\begin{align*}
    E_k(\mu) &= P(\text{Accept } H_0 \text{ at stage } k | \mu),\\
    F_k(\mu) &= P(\text{Reject } H_0 \text{ at stage } k | \mu),\\
    S_k(\mu) &= P(\text{Trial terminates after stage } k | \mu),\\
             &= E_k(\mu) + F_k(\mu).
\end{align*}

Then, the ESS for a GSD is often stated as

\begin{align*}
 ESS(\mu) &= \sum_{k = 1}^K \{E_k(\mu) + F_k(\mu)\}n_k,\\
          &= \sum_{k = 1}^K S_k(\mu)n_k.   
\end{align*}

Therefore, the expected efficiency gain (EG) from using a GSD instead of a corresponding single-stage design can be calculated as

$$ EG(\mu) = \frac{n_\text{single} - ESS(\mu)}{n_\text{single}}, $$

where $n_{single}$ is the required sample size for the single-stage design.

\subsection{Efficiency accounting for outcome delay}

The formula above for the ESS ignores the potential issue of outcome delay (i.e., it essentially assumes that outcome $X_{ij}$ is accrued immediately after recruitment).
To extend the formulae above to allow for outcome delay, we suppose that responses are available a time $m$ after a participant is recruited.
If we assume that recruitment is not paused for the conduct of each IA, there will then be additional participants recruited between the recruitment of participant $n_k$ and the conduct of IA $k$ (these are often referred to as `pipeline' participants).
We will denote this random variable, i.e., the number of such pipeline participants at the time of IA $k$ by $\tilde{n}_k$ for $k = 1, 2, \dots, K - 1$.
We assume that recruitment stops when $n_K$ participants have been recruited, such that there can be no pipeline participants at analysis $K$.

To quantify the efficiency lost due to delay, we therefore require expected values for the $\tilde{n}_k$.
These values will depend on the delay length $m$, but also on the recruitment model.
They will be the focus of the coming sections, where we define a framework in which recruitment will be a function of parameters $\delta$, $l$, and $t_\text{max}$, to be defined later.
Accordingly, we have $\tilde{n}_k = \tilde{n}_k(m, \delta, l, t_\text{max})$ and the ESS when accounting for outcome delay will then be given by
\begin{small}
\begin{align*}
   ESS_\text{delay}(\mu, m, \delta, l, t_\text{max}) &= \sum_{k = 1}^{K - 1} \{E_k(\mu) + F_k(\mu)\}(n_k + \tilde{n}_k) \\
                                                     & \qquad \qquad + \{E_K(\mu) + F_K(\mu)\}n_K, \\
                                                     &= \sum_{k = 1}^{K - 1} S_k(\mu)\left(n_k + \tilde{n}_k\right) + S_K(\mu)\ n_K.
\end{align*}
\end{small}

Thus, the `true' efficiency gained (EG) compared to a single-stage design in the presence of outcome delay will be measured as
\begin{small}
$$ EG_\text{Delay}(\mu, m, \delta, l, t_\text{max}) = \frac{n_\text{single} - ESS_\text{Delay}(\mu, m, \delta, l, t_\text{max})}{n_\text{single}}. $$
\end{small}
We will then quantify the efficiency loss (EL) due to outcome delay as the percentage change in the EG when considering delay in comparison to not considering delay.
That is
\begin{small}
$$ EL(\mu, m, \delta, l, t_\text{max}) = 100\frac{EG(\mu)-EG_\text{delay}(\mu, m, \delta, l, t_\text{max})}{EG(\mu)}. $$
\end{small}

The value of EL is typically expected to be less than or equal to 100, however,it is possible to observe a value greater than 100 if the delay makes the group sequential design less efficient than the single stage design.

\subsection{Computing the number of pipeline participants}

We consider two sub-cases for estimating the number of pipeline participants at a given IA.
From here onward, we consider time to be a discrete variable, as previous work indicates minimal difference from treating time as continuous. 
Further, GSDs usually tend to be long trials, which makes the formulae simpler to communicate and comprehend, when time is assumed to be discrete. 
We also assume the unit of time to be months.
The results could also be readily generalised for other units of time, given all the parameters are defined in the same units.

\subsubsection{Uniform recruitment}

We consider a uniform recruitment pattern with rate of recruitment $\lambda$.
We suppose it takes $t_\text{max}$ months to recruit all $n_K$ participants.
Then, for uniform recruitment, the expected number of pipeline participants at each IA should typically be constant, say $\tilde{n}$.
However, we must account for the fact that the number of pipeline participants cannot lead to the total sample size of the trial being above $n_K$.
Thus, in this case
\begin{align*}
    \tilde{n}_k &= \begin{cases}
                   \tilde{n} &: \tilde{n} \le n_K - n_k,\\
                   n_K - n_k &: \tilde{n} > n_K - n_k,
                   \end{cases}
\end{align*}
where
\begin{align*}
    \tilde{n} = \lambda m = \frac{n_K}{t_\text{max}}m.
\end{align*}

\subsubsection{`Mixed' recruitment}

\begin{figure}
    \centering
    \includegraphics[width=0.48\textwidth]{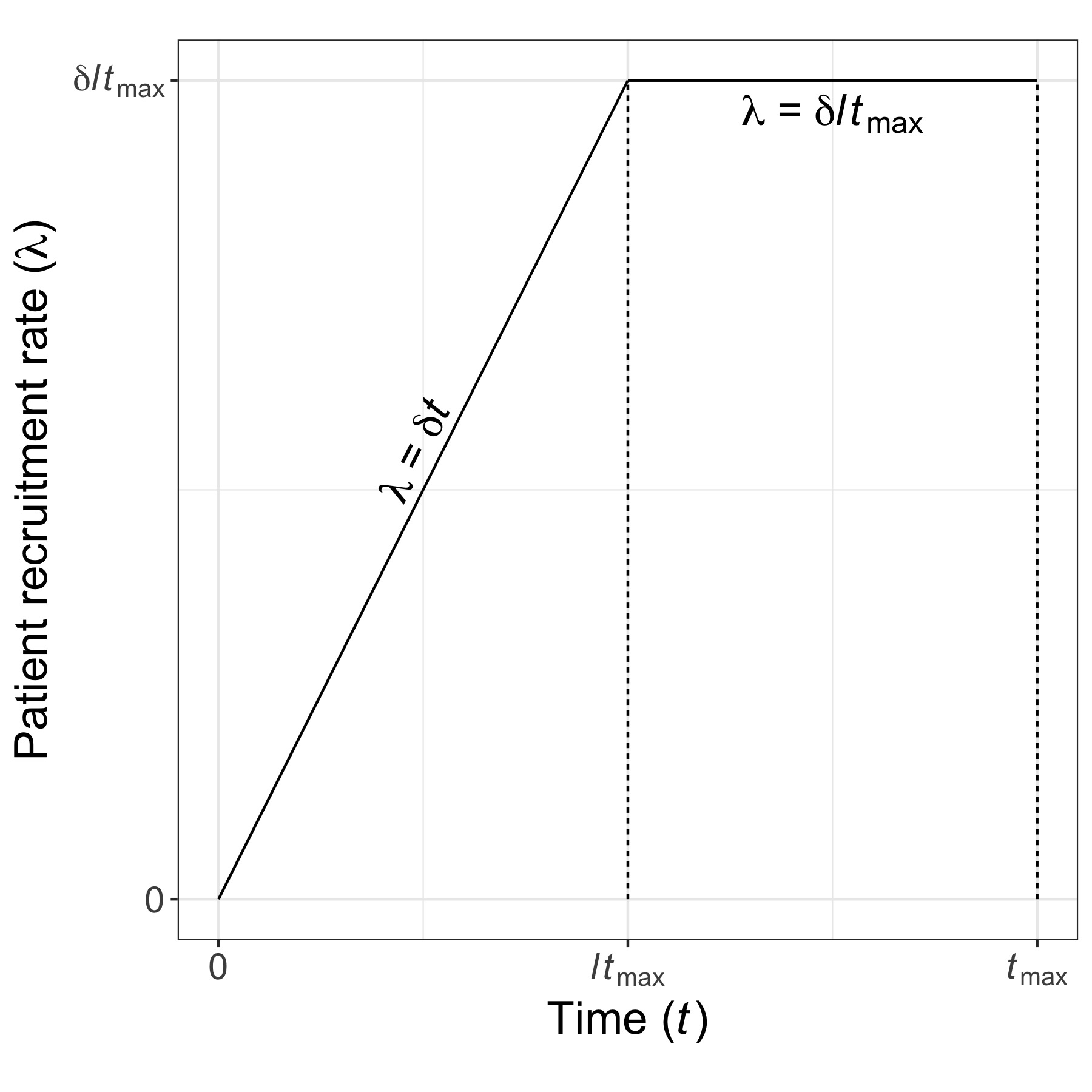}
    \caption{Recruitment model for the `mixed' recruitment pattern.\label{Fig1}}
\end{figure}

In reality, uniform participant recruitment may poorly reflect recruitment rates observed in practice.
This is because early in a trial, sites are gradually opened until a maximum number is reached.
We allow for this by assuming participants are recruited at time $t$ in a linearly increasing pattern (at rate $\lambda = \delta t$) up to $l$ times of the total recruitment length $t_\text{max}, 0 < l \le 1$ (see Figure~\ref{Fig1}).
For times above $lt_\text{max}$, we assume the recruitment pattern is then uniform, with $\lambda = \delta lt_\text{max}$.
We refer to this more general pattern of recruitment as `mixed recruitment'.
Note that when $l = 1$ we observe a continuously linearly increasing recruitment pattern; we will refer to this special case later as `linear recruitment'.
We assume throughout that (assumed) values for $l$ and $t_\text{max}$ have been specified, reflecting the common practice at the design stage of any study in which recruitment must be projected.

Next, denote by $t_k$ the expected amount of time taken to recruit $n_k$ participants.
Then, $\tilde{n}_k$ will depend on $t_k$, $m$, $\delta$, and $l$.

Observe that under the above recruitment model, in $t_\text{max}$ months the total number of recruitments is expected to be
\begin{align*}
    & \delta(1 + 2 + \cdots + lt_\text{max}) + \delta lt_\text{max} (1 - l) t_\text{max}\\
    & = 0.5\delta lt_\text{max} (lt_\text{max} + 1) + \delta lt_\text{max} (1 - l) t_\text{max}.
\end{align*}
As this value should equal the maximum sample size $n_K$, this provides us with an estimate for $\delta$, as the other quantities in the above formula are fixed.

To compute general estimates for the $\tilde{n}_k$, we must account for several possibilities, based on the location of $lt_{max}$ in the recruitment rate.

\begin{itemize}
    \item If $lt_\text{max} < t_1$, i.e., the first IA happens after the recruitment rate becomes uniform, then the expected number of pipeline participants at each IA remains constant due to the uniform recruitment pattern and takes the value $\tilde{n}_k = \delta lt_\text{max}m$.\\

    \item When $lt_\text{max}$ lies between IA $i$ and $i + 1$, i.e., $t_i \le lt_\text{max} < t_{i + 1}$ for $i= 1, 2,\dots, K-2$, then $\tilde{n}_k = \delta lt_\text{max}m$ for $k = (i+1), (i+2), \dots, (K - 1)$, due to the uniform recruitment pattern. (Note that for $i = K - 1$, there would be no pipeline participants at analysis $i + 1 = K$ as this is the final stage of the trial.)\\

    For the expected number of pipeline subjects for IA $k = 1, 2, \dots, i$, we need the values of $t_k$, $k = 1, 2, \dots, i$ as the recruitment is assumed to be linear for IA $k = 1, 2, \dots, i$.\\
    It can be computed as
    \begin{align*}
        \delta(1+2+\dots+t_k)=n_k\\
    or, \delta \frac{t_k(t_k+1)}{2}=n_k
    \end{align*}
    Solving the quadratic in $t_k$ gives
    \begin{align*}
     t_k=-0.5+0.5\sqrt{(1+\frac{4*2n_k}{\delta})}
    \end{align*}

    Then, the expected number of pipeline participants for IA $k, k= 1, 2,\dots, (i-1)$ is given by 
    
      \begin{align*}
            \tilde{n}_k &= \delta\{(t_k + 1) + (t_k + 2) + \cdots + (t_k + m)\},\\
                        &= \delta mt_k + \delta m(m + 1)/2.
        \end{align*}

For $k = i$, the value of $\tilde{n}_k$ depends on the location of $lt_\text{max}$ as follows
    
    \begin{enumerate}
        \item If $t_i + m < lt_\text{max}$, then the pipeline participants are obtained from assuming linearly increasing recruitment as in the previous case,
        \begin{align*}
            \tilde{n}_i &= \delta\{(t_i + 1) + (t_i + 2) + \cdots + (t_i + m)\}, \\
                        &= \delta m t_i + \delta m(m + 1)/2.
        \end{align*}
        \item If $t_i + m \ge lt_\text{max}$ then the pipeline participants are obtained from assuming linearly recruitment first and uniform recruitment for the remaining time. This gives
        \begin{align*}
            \tilde{n}_i &= \delta\{(t_i + 1) + (t_i + 2) + \dots + lt_\text{max}\} \\ 
                        & \qquad + \delta lt_\text{max}(t_i + m - lt_\text{max}).
        \end{align*}
    \end{enumerate}
\end{itemize}

Therefore, in summary, for  $i= 1, 2, \dots, K-2$, the number of pipeline subjects can be found as in table~\ref{Pipelines}.

\begin{table*}[h]
\centering
\caption{Number of Pipeline subjects for $t_i \le lt_\text{max} < t_{i + 1}$, $i= 1, 2, \dots, K-2$ }
\label{Pipelines}
\begin{tabular}{@{}lll@{}}
\bottomrule
IA & Position of $lt_{max}$ & Estimate of pipeline subjects \\ \bottomrule
$k=1,2,\dots, (i-1)$ &  & $\tilde{n}_k = \delta mt_k + \delta m(m + 1)/2$ \\
 &  &  \\
$k=i$ & $t_i + m < lt_\text{max}$ & $\tilde{n}_i =  \delta m t_i + \delta m(m +   1)/2$ \\
      & $t_i + m \ge lt_\text{max}$ & $\tilde{n}_i= \delta\{(t_i + 1) + (t_i + 2) + \dots +lt_\text{max}\}+$\\
 &  &    $  \delta lt_\text{max}(t_i + m -   lt_\text{max}).$\\
 &  &  \\
$k=(i+1),(i+2),\dots, K-1$ &  & $\tilde{n}_{k} = \delta lt_\text{max}m$ \\ 
\bottomrule
\end{tabular}
\end{table*}

\subsection{Impact of delay on expected time to trial completion}

So far, we have considered the ESS as the measure of efficiency for analysing the impact of outcome delay on the trial design.
However, particularly in clinical trials sponsored by the pharmaceutical industry, the primary measure of optimality, might not be a reduced ESS but the overall time to complete the trial.
In this section, we therefore explore how a delay in observing the treatment outcome impacts the expected time to complete the trial.

Let us denote by $T$ the time to complete a trial for a $K$ stage design.
Now, at IA $k$, $T$ is given by the sum of the total time to recruit stage $k$ participants (say, $t_k$) and time to observe their treatment outcome$(m, say)$; $k=1,2,\dots,K$\\

Let us assume for a group-sequential design with $K$ stages, the sample size at stage $k$ is given by $n_k, k=1,2,\dots,K.$ Suppose, the participants are being recruited uniformly over a total recruitment period of $t_\text{max}$.

Then the time taken to recruit all the participants $(n_k)$ for stage k would be given as
$$ t_k = \frac{t_\text{max}}{n_K}n_k,\ k = 1, 2, \dots, K. $$

If we consider the delay into account then, the expected time to complete the trial can be given as

\begin{align*}
   ET(\mu) &= \sum_{k = 1}^K (t_k + m)S_k(\mu) \\
           &= m + \sum_{k = 1}^K t_kS_k(\mu) \\
           &= m + \sum_{k = 1}^K \frac{t_\text{max}}{n_K}n_kS_k(\mu) \\ 
           &= m + \frac{t_\text{max}}{n_K}ESS(\mu).
\end{align*}

Note that if we consider that the IA will take some time to conduct, say $m_{interim}$ months then, this quantity will only add up in the above equation, i.e. $ET(\mu)$ becomes$ [m + m_{interim}  +t_{max}/n_K  ESS(\mu)]$.

Therefore, under the assumption that the recruitment is uniform the expected time to complete a trial is a linear function of the ESS.

For a linearly increasing recruitment pattern, $t_k$ can be obtained by solving the following equation
\begin{align*}
   & \delta(1 + 2 + \cdots + t_k) = n_k \\
    \Rightarrow & \frac{\delta t_k(t_k + 1)}{2} = n_k \\
    \Rightarrow & t_k = \frac{-1 + \sqrt{1 + \frac{8n_k}{\delta}}}{2}.
\end{align*}
where,
$$ \delta = \frac{2n_K}{t_\text{max}(t_\text{max} + 1)}. $$

If we assume the same recruitment rate as before (i.e. $t_{max}/n_K$ ), the expected time to complete a traditional RCT is given by $T = t_{single}+m$, where, $t_{single}  =n_{single}  \frac{t_{max}}{n_K}$  is the time required to recruit the total of $n_{single}$ samples required for the traditional RCT.
The above results show that, given a total recruitment period, $ET (\mu)$ is a linear function of ESS or a function of lower degree of the sample size at each stage. 
We know that a GSD without delay typically provides a benefit in terms of lower ESS, compared to a traditional RCT, especially when the alternative is true. 
Since the $ET (\mu)$ is a linear function of the ESS, or a function of the sample size of lower degree, $ET (\mu)$ will then also typically be lower than that of a traditional RCT design, assuming a similar recruitment rate.

Therefore, it is generally beneficial on average to conduct a GSD if the efficiency metric is the time to complete a trial.

\subsection{Examples}

Although the expected time to complete the trial gives definitive answers when a group sequential is beneficial, the efficiency gain in terms of ESS in presence of delay still remains under investigation.
Therefore the following section provides a roadmap to different simulation scenario for assessing the impact of delay on group sequential trials.

For our study, we have considered both equally and unequally spaced IAs with uniform, linear, and mixed recruitment patterns.
In practice, most trials using a GSD have a maximum of $K = 5$ stages \cite{stevely2015}.
Therefore, we focused our results on designs with $K = 2$, 3, 4, or 5.

In order to allow for futility stopping, all the designs considered onwards use futility stopping boundaries of 0. 
Thus, the stopping boundaries are asymmetric. 
However, results for symmetric boundaries are provided in the section \ref{section 3.3}. and in more details in the Supplementary Materials.

Throughout, we set $\alpha = 0.05$, $\beta = 0.1$, and $\mu = \tau = 0.5$ (i.e., the EL is evaluated under the target effect).
Also, we assume equal allocation to the  experimental and control arms (i.e., $n_{0k} = n_{1k}$ for $k = 1, 2, \dots, K$) and $\sigma_0^2 = \sigma_1^2 = 1$.

The total recruitment period is assumed to be $t_\text{max} = 24$ months and we provide results for varying delay periods up to 24 months.
(Exact EL values are provided for delay length of $m = 3$, 6, 9, 12, 18, and 24 months in the supplementary materials.)
For the mixed recruitment pattern, in all of the above cases we considered scenarios when $l = 0.2$, 0.4, 0.6, 0.8, i.e., the trials had a linear recruitment rate for 20, 40, 60, or 80\% of the total recruitment period.

For unequally spaced IAs, we have considered four different combinations of IA spacings in the 3 stage design setting.
Specifically, we assumed the information fractions $(\rho_1, \rho_2, \rho_3)$ to be $(1/3, 2/3, 1)$, $(1/4, 1/2, 1)$, $(1/2, 3/4, 1)$, or $(6/10, 9/10, 1)$.
For 4 stage designs, we similarly considered different combinations for IA spacing viz. $(\rho_1, \rho_2, \rho_3, \rho_4)$ to be $(1/4, 2/4, 3/4, 1)$, $(1/5, 2/5, 3/5, 1)$, $(2/5, 3/5, 4/5, 1)$

R code to reproduce our analyses is available from \url{https://github.com/AritraMukherjee/GroupSequentialDesign.git.}

\section{Results: Impact of delay on expected sample size}

First, we quantify the efficiency lost in terms of ESS as a result of delay in observing treatment outcomes.
We subset our results by the spacing of the IAs.

\subsection{Equally spaced interim analyses}
 
The following subsection contains results assuming GSDs with Wang-Tsiatis boundaries \cite{wang1987}; the value of the shape parameter is assumed to be 0.25.
Results for other boundary shapes (e.g., Pocock, O'Brien-Fleming) can be found in the Supplementary Material.

\begin{figure}
    \centering
    \includegraphics[width=0.6\textwidth]{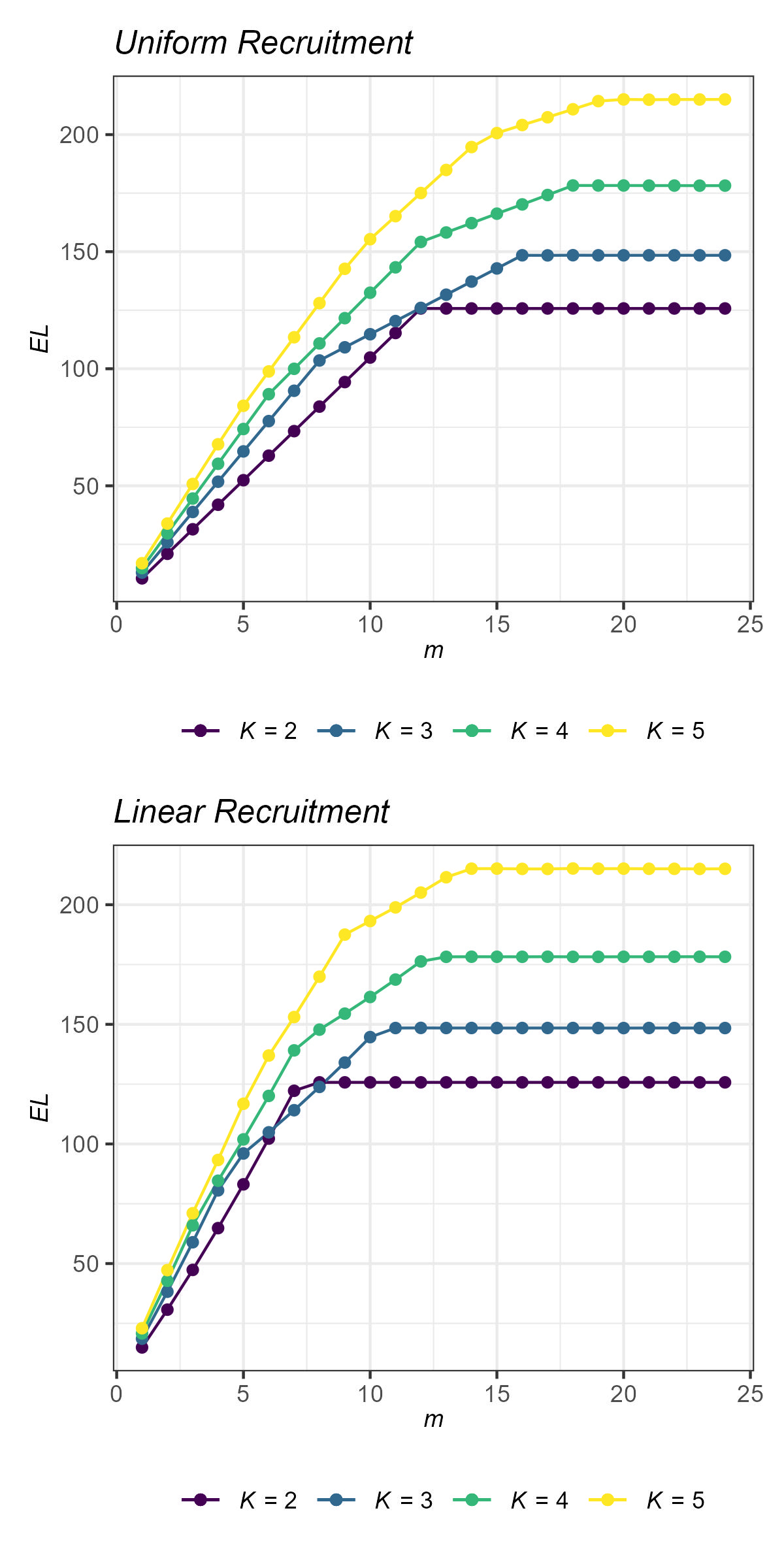}
    \caption{Efficiency loss (EL) due to delay, for different delay lengths $m$, assuming equally spaced interim analyses, under uniform and linear recruitment patterns.\label{Fig2}}
\end{figure}

\begin{figure}
    \centering
    \includegraphics[width=0.6\textwidth]{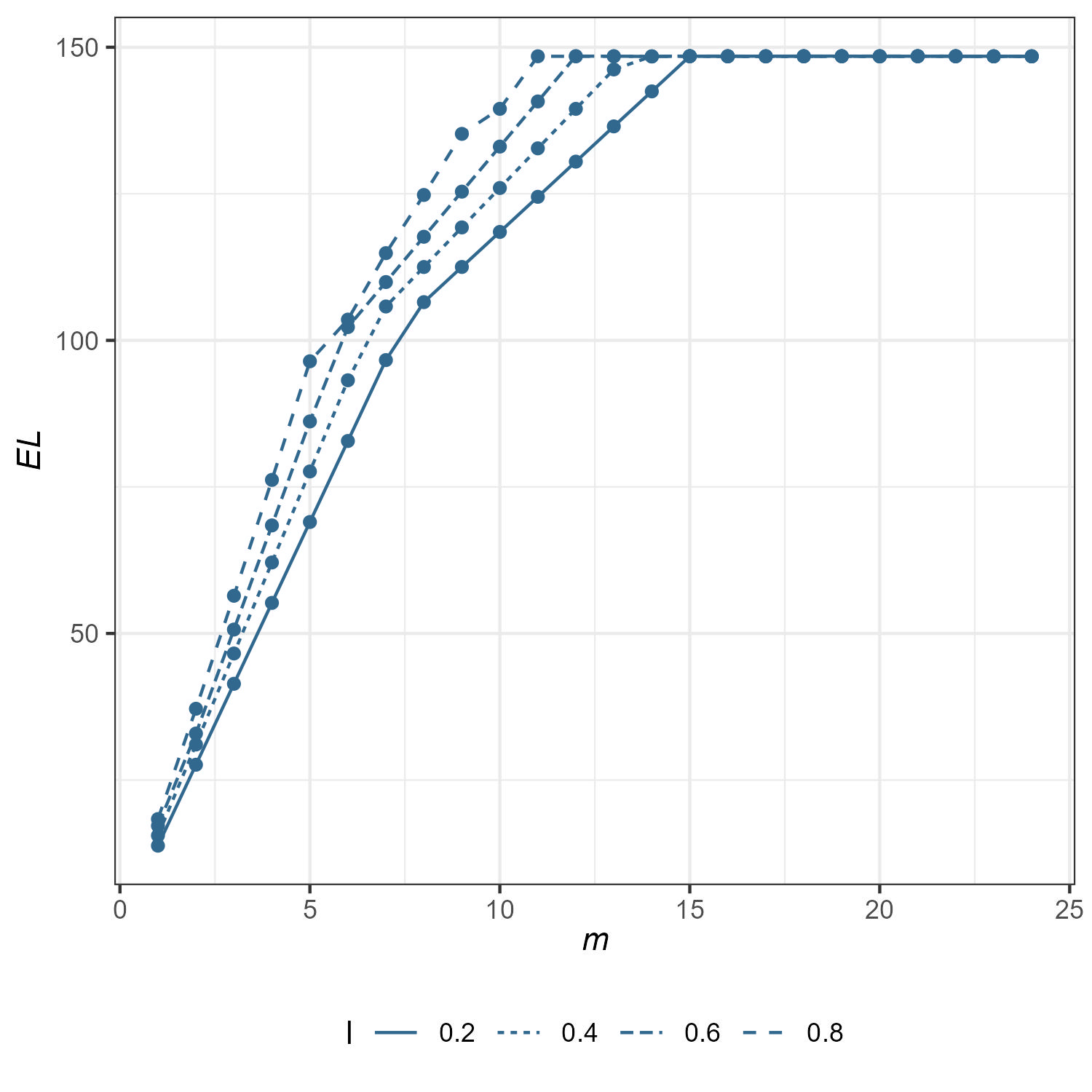}
    \caption{Efficiency loss (EL) due to delay, for different delay lengths $m$, assuming equally spaced interim analyses in a 3-stage design, under a mixed recruitment pattern.\label{Fig3}}
\end{figure}

From Figure~\ref{Fig2}, it can be observed that as $m$ increases, there is an increasing EL due to delay.
An EL of 100\%, indicates that the EG expected for using a GSD is completely lost due to delay, i.e. the value of $ESS_\text{delay}$ is same as the single stage sample size.
For higher EL values, the $ESS_\text{delay}$ is even greater than $n_\text{single}$, where $ESS_\text{delay}$ attains the maximum possible sample size for the design. 

A linearly increasing recruitment pattern incurs heavy efficiency loss, when compared to a uniform recruitment pattern, even for smaller delay lengths, due to the recruitment pattern. 
The EL attains a distinct, maximum value for each $K$ as $ESS_{delay}$ attains a constant maximum value (the maximum sample size) after a certain delay length, based on the design. 
The EL appears to increase with the number of stages. 
For this study, we have assumed a constant recruitment period of 24 months for all the designs, and the maximum sample size for different number of stages varies significantly. 
As a result, the recruitment rate increases marginally for designs with increasing $K$. 
Therefore, with a higher $K$ (and thus a higher recruitment rate) the number of pipelines accumulated increases Hence, as $K$ increases, a small increase in the EL value for the same $m$ occurs. 
It can be inferred that, by a small amount, a 2-stage design can be beneficial in terms of a reduced EL as compared to a $K$-stage design, where, $K\ge 3$. 
Furthermore, for $m$ greater than 4 months, almost 50\% of the expected EG is typically lost due to delay for all the GSDs.

For almost all values of $K$, the minimum value $m$ needed for a GSD to attain its maximal EL is 15 to 18 months for uniform recruitment. 
While for linear recruitment the maximum EL is attained even sooner, at approximately 12 months delay.

For uniform recruitment with small delay $(m = 2)$, the maximum loss observed is 33.87\% for $K = 5$ and the minimum is 20.95\% for $K = 2$. 
The same values for linearly increasing recruitment are 47.27\% for $K = 5$ and 30.72\% for $K = 2$ respectively. 
Therefore, for smaller delay lengths, GSDs retain most of their efficiency gain. 
On the other hand, the maximum EL observed is 215\% for a 5-stage design when the delay length is greater than 14 months, under both uniform and linear recruitment, i.e. the trial ends up recruiting the maximum sample size which is twice the number of the ESS initially planned for.

For the mixed recruitment pattern, we provide results for $l = 0.2,0.4,0.6,$ and 0.8. Findings for a 3-stage design are shown in Figure \ref{Fig3}. It can be observed that the results obtained align with the findings in Figure \ref{Fig2}, i.e., as the recruitment pattern becomes linear for a greater proportion of the total recruitment time, the EL increases. 
Further computations also indicate that, as the number of stages increases the EL is increased by a small amount (see Supplementary Materials).

In summary, it was observed that if the delay length is more than 25\% of the total recruitment period, at least 50\% of the expected EG is lost due to delay for all recruitment patterns for 2-5 stage designs.

\subsection{Unequally spaced interim analyses}

For three-stage designs, we considered four possible timings of the IAs under uniform and linear recruitment patterns.
These were

\begin{figure}
    \centering
    \includegraphics[width=0.6\textwidth]{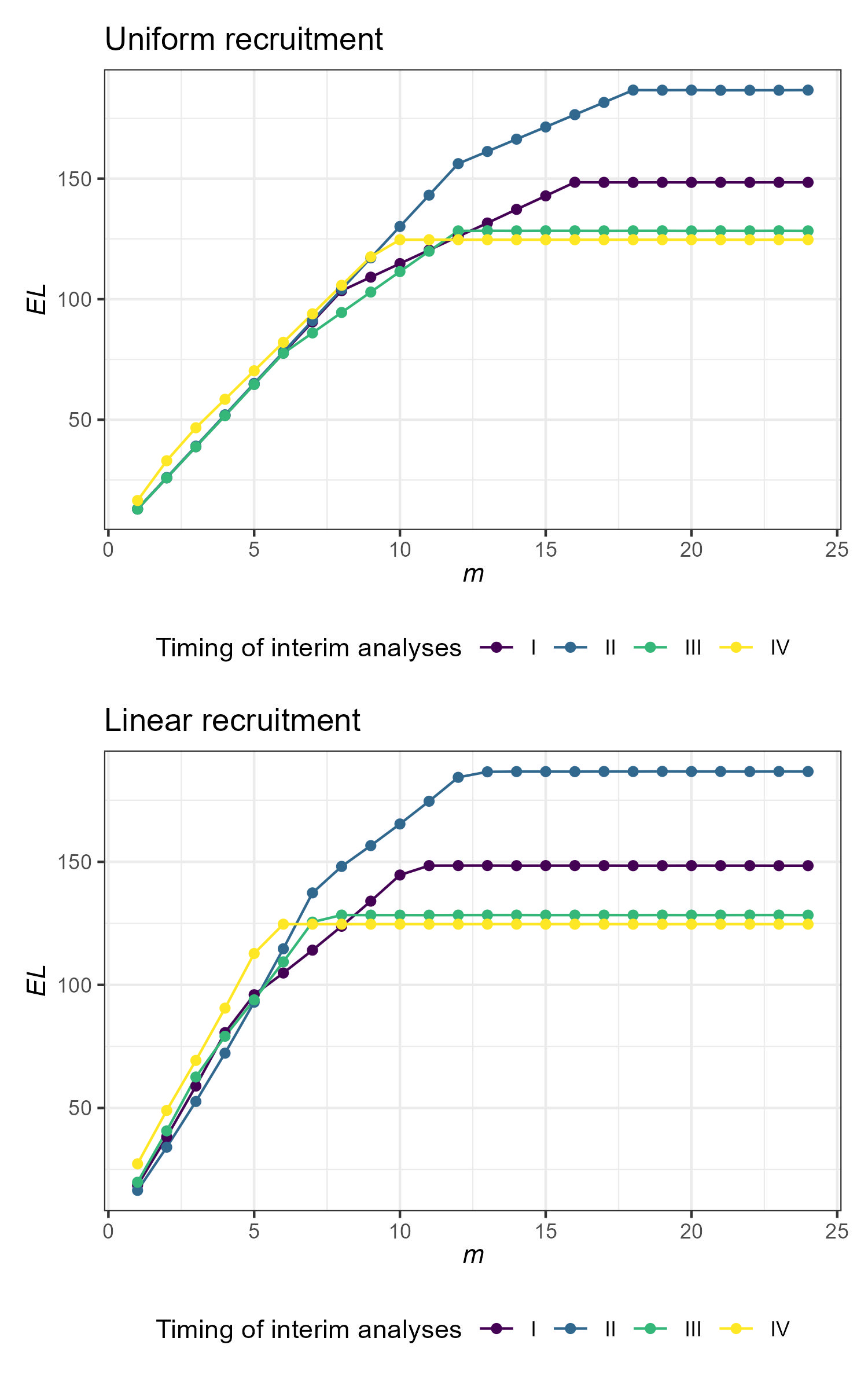}
    \caption{Efficiency loss (EL) due to delay for different delay lengths, in 3-stage designs with unequally spaced interim analyses, under uniform and linear recruitment patterns.\label{Fig4}}
\end{figure}

\begin{enumerate}
    \item $(1/3, 2/3, 1)$: IAs are equally spaced (I).
    \item $(0.25, 0.5, 1)$: The IAs takes place sooner than under equal spacing (II).
    \item $(0.5, 0.75, 1)$: The IAs take place later than under equal spacing (III).
    \item $(0.6, 0.9, 1)$: The IAs occur even later than the design above (IV).
\end{enumerate}
Figure~\ref{Fig4} shows the results for the above scenarios.

It can be observed that, in general, the different designs with different interim spacings undergo a similar EL for small delay lengths, especially for uniform recruitment. 
This is due to a similar number of pipelines being recruited in the trial as the recruitment rate observed for different interim spacings remains very similar. 
The ESS for lower values of m also take values very close to each other. Therefore, the loss typically takes very similar values for $m\le 8$. 
However, as $m$ increases further, it is observed that pushing the IA towards the latter half of the trial can prove to be beneficial for the design. 
The case when the first IA is performed even sooner than that under equal spacing, i.e., the design where the two IAs are conducted at 25\% and 50\% of the total sample size incurs the greatest losses. 
This is because for designs with IAs occurring towards the latter end of the trial, the maximum sample size is relatively lower than an equally spaced design. 
Therefore, the number of pipelines that can be recruited in the trial is restricted to a lower value as compared to an equally spaced design (interim spacing I) or a design where the interim analysis is performed sooner (interim spacing II) than that of an equally spaced design. 
For a linearly increasing recruitment the maximum loss is observed much sooner due to the increased number of pipelines at an earlier time point. 

The maximum EL observed for IA timings (0.6, 0.9, 1) is 125\%, in contrast to 187\% for IA timing (0.25, 0.5, 1). 
This EL occurs much sooner at a 6-month delay, rather than 10 months, under linear recruitment.

We also considered four-staged GSDs, under unequally spaced interims for different combinations of interim spacings.
The results obtained are very similar to those for a three-stage design, i.e., if the first IA is pushed towards the latter end of the trial, the EL is decreased when the delay length is sufficiently large ( $\ge 7$ months).
See the Supplementary Materials for these findings. 

It is interesting to note that, if the recruitment period is fixed in advance, further computations also revealed that for any other choice of the recruitment period(e.g. 36 months), similar efficiency losses would be observed for similar values of $\frac{m}{t_{max}}$. i.e. the EL value remains similar for a 18 month delay for 36 month recruitment and a 12 month delay for 24 month recruitment. 
Therefore, the inferences can be generalised in terms of the ratio $\frac{m}{t_{max}}$.

\subsection{Symmetric stopping boundaries}\label{section 3.3}
\begin{figure}
    \centering
    \includegraphics[width=0.5\linewidth]{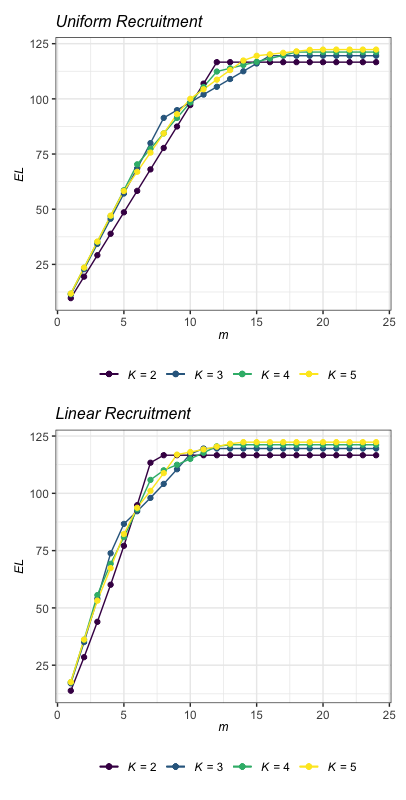}
    \caption{Efficiency loss (EL) due to delay, for different delay lengths $m$, assuming equally spaced interim analyses in group-sequential designs with different numbers of stages $K$, under uniform and linear recruitment patterns. Here, a symmetric WT stopping boundary has been used.}
    \label{fig5}
\end{figure}

The results so far describe the impact of delayed outcomes for asymmetric stopping boundaries. 
In this section, we consider symmetric stopping boundaries for IAs, i.e. for $k=1,2,…,K-1;f_k=-e_k$  and $f_K=e_K$. 
The efficacy and futility boundaries were chosen assuming the Wang-Tsiatis boundary with shape parameter 0.25. 
Figure ~\ref{fig5} plots the EL assuming an equally spaced GSD for $m=1,2,…,24$ and $K=2,3,4,5$.

Similarly to Figure \ref{Fig2}, Figure \ref{fig5} also shows that, with an increase in $m$, EL increases and a linearly increasing recruitment pattern incurs a heavier efficiency loss than uniform recruitment. 
However, the EL value is much lower compared to an asymmetric stopping boundary. 
This can be mainly attributed due to a lower stopping probability at earlier stages of the design (the probability of stopping for futility is almost 0). 
Furthermore, the EL appears to be similar for different number of stages, especially for $K = 3,4$ and 5 due to the stopping probabilities. 
In addition, for a GSD with symmetric WT stopping boundary, the maximum sample size for different number of stages varies by only a small amount. 
Therefore, the recruitment rate remains similar for designs with different $K$ under a constant recruitment period of 24 months.
Hence, the number of pipelines accumulated for different values of $K$ takes close values especially for uniform recruitment. 
As a result, a similar EL can be observed for varying $K$ [especially K=3,4 and 5]. 
The EL attains a close, but distinct, maximum value for each $K$ as $ESS_{delay}$ attains a constant maximum value (the maximum sample size), based on the design. 

For almost all values of $K$, the minimum value $m$ needed for a GSD to attain its maximal EL is 15 months for uniform recruitment. 
For linear recruitment the maximum EL is attained even sooner, at approximately 12 months delay.

We observed similar patterns for EL for mixed recruitment patterns for symmetric boundaries and asymmetric boundaries (see Supplementary Materials for the results), i.e. as the recruitment became more linear, EL was increased. 
More details on the impact of delay on designs with symmetric boundaries are discussed in the Supplementary Materials.

\section{Case Study}

To supplement previous results and to assess the efficiency lost in a real trial, we consider a trial that assessed the efficacy of agomelatine combined with antipsychotics to treat negative symptoms in schizophrenia (NCT05646264). 
This was a prospective, multicenter, controlled trial that used a GSD to allow early stopping for efficacy and futility. 
The primary outcome was the reduction in total PANSS negative scale score at 24 weeks. 
The treatment regimen was to be considered effective if this rate of reduction was greater than or equal to 20\%. 
The trial specified a target sample size of 220 participants to be recruited in 7 months, 2023-01-01 to 2023-08-01. 
Therefore, for this example, $m = 6$ with $t_{max}  = 7$.

As the specific details of the GSD (e.g., number of stages) are not provided on the clinicaltrials.gov registration page, we provide expected ELs assuming different number of stages and stopping boundaries. 
We assume, the sample size was computed based on 5\% significance level and 90\% power, as this results in a similar value to that listed as required.
We consider Pocock, O’Brien-Fleming (OBF), and Wang-Tsiatis (WT) boundaries for obtaining the results (OBF stopping boundaries for K = 3 give the closest sample size to that required).

Our findings are given in Table \ref{table2}. 
It can be seen that, for the given trial, the GSD fails to provide any advantage in terms of a reduced ESS, due to the delay in observing the treatment outcome. 
If we consider the case where the sample size matches the design mentioned in the study, i.e., a 3-stage design with OBF stopping rules, the expected EL is 111\%. 
In this example, by the end of the first IA, given the recruitment rate, the trial would have recruited another 146 participants, leading to a total recruited sample size of 220 at the end of first stage. 
Therefore, $ESS_{delay}$ takes the value of the maximum sample size 220. 
By contrast, a single stage design for the same power and significance level would have required only 214 participants. 
For Pocock and WT stopping boundaries, the EL increases, also making the single stage design a better choice.
Note that Table 2 provides the number of pipeline participants assuming uniform recruitment. 
However, due to the large delay, the loss would be the same under linear recruitment.

\begin{landscape}
\begin{table}[]
\centering
\caption{Efficiency lost due to delay for the considered case study (NCT05646264). The EL is provided for different stopping boundaries and numbers of stages, alongside the stage wise sample sizes and number of pipeline participants recruited during the 6-month delay period}
\label{table2}
\begin{tabular}{@{}ccccccccccccccc@{}}
\toprule
\textbf{Stopping} &
  \textbf{$K$} &
  \textbf{} &
  \textbf{} &
  \textbf{$n_k$} &
  \textbf{} &
  \textbf{} &
  \textbf{} &
  \textbf{} &
  \textbf{$\tilde{n}_k$} &
  \textbf{} &
  \textbf{} &
  \textbf{$ESS$} &
  \textbf{$ESS_{delay}$} &
  \textbf{$EL$} \\ \cmidrule(lr){3-12}
\textbf{Boundary} &   & $k=1$ & $k=2$ & $k=3$ & $k=4$ & $k=5$ & $k=1$  & $k=2$  & $k=3$  & $k=4$ & $k=5$ &        &        &        \\ \midrule
Pocock            & 2 & 119   & 238   &       &       &       & 118.78 & 0      &        &       &       & 163.96 & 237.55 & 146.77 \\
                  & 3 & 84    & 167   & 250   &       &       & 166.36 & 83.18  & 0      &       &       & 151.25 & 249.54 & 156.39 \\
                  & 4 & 65    & 129   & 193   & 258   &       & 192.98 & 128.66 & 64.33  & 0     &       & 145.82 & 257.31 & 163.30 \\
                  & 5 & 53    & 106   & 158   & 211   & 263   & 210.37 & 157.78 & 105.19 & 52.59 & 0     & 142.94 & 262.97 & 168.68 \\
                  \bottomrule
OBF               & 2 & 109   & 218   &       &       &       & 108.57 & 0      &        &       &       & 175.2  & 217.14 & 107.83 \\
 &
  \textbf{3} &
  \textbf{74} &
  \textbf{147} &
  \textbf{220} &
   &
   &
  \textbf{146.28} &
  \textbf{73.14} &
  \textbf{0} &
   &
   &
  \textbf{165.66} &
  \textbf{219.42} &
  \textbf{110.00} \\
 &
  \textbf{4} &
  \textbf{56} &
  \textbf{111} &
  \textbf{166} &
  \textbf{221} &
   &
  \textbf{165.73} &
  \textbf{110.48} &
  \textbf{55.24} &
  \textbf{0} &
   &
  \textbf{159.52} &
  \textbf{220.97} &
  \textbf{112.59} \\
                  & 5 & 45    & 89    & 134   & 178   & 223   & 177.63 & 133.22 & 88.82  & 44.41 & 0     & 155.86 & 222.04 & 113.64 \\
                  \bottomrule
WT                & 2 & 113   & 225   &       &       &       & 112.03 & 0      &        &       &       & 166.29 & 224.06 & 120.84 \\
$(\Delta=0.25)$   & 3 & 77    & 153   & 229   &       &       & 152.13 & 76.07  & 0      &       &       & 155.13 & 228.2  & 123.91 \\
                  & 4 & 58    & 116   & 173   & 231   &       & 172.96 & 115.31 & 57.65  & 0     &       & 149.70  & 230.61 & 125.65 \\
                  & 5 & 47    & 93    & 140   & 186   & 233   & 185.79 & 139.35 & 92.9   & 46.45 & 0     & 146.41 & 232.24 & 126.81 \\ 
                  \bottomrule
\end{tabular}
\end{table}
\end{landscape}

\section{Discussion}

GSDs have been both widely used in practice and explored methodologically.
However, little work has considered the impact of the time taken to observe the primary outcome variable when examining the utility of a GSD.
This despite outcome delay being harmful to the efficiency of a GSD.
Our study therefore aimed to explore the extent to which group-sequential trials could be impacted by outcome delay.
An EL was computed based on the difference in the efficiency gained over a single-stage trial without delay and with delay.
We estimated the number of pipeline participants assuming uniform recruitment, linearly increasing recruitment, and under `mixed' recruitment which combined these two patterns.
Uniform recruitment is more likely a reasonable assumption for smaller scale single-centre trials, whereas mixed recruitment is more reasonable for large multi-centre trials, where the recruitment rate may increase as new centres open and reach a maximum rate once all centres are operational.
A linearly increasing recruitment pattern can then be considered as an extreme case, where the recruitment rate never plateaus during the length of the enrollment.

We obtained results for different recruitment patterns and different delay lengths.
They showed that, as would be expected, with an increase in the delay length the EL increases; this is a consequence of the fact that as delay increases so does the number of overruns, thereby increasing $ESS_\text{delay}(\cdot)$.
The EL also increases with different values of $K$. 
Significantly, it was observed that when $\frac{m}{t_{max}}$  takes values more than 0.5, GSDs incur heavy losses due to delay.

An important point to note is that tables given in the Supplementary Materials indicate that, there are times when $ESS_{delay}$ for a 2-stage design still remains greater than $ESS_{delay}$ for a 3-stage design, even in situations where the EL is smaller for the 2-stage design. 
That is, it might be beneficial to use a 3-stage design since the $ESS_{delay}$ value is lower, even though it suffers a greater EL.
However, for large delay lengths, designs with more IAs can lose efficiency more quickly. 
Thus, with large delay present, it may be better to reduce the number of IAs to reduce the impact of delay. 
In all, careful inspections are necessary to select the best possible design.

We also observed that the EL is typically greater under a linearly increasing recruitment pattern than for uniform recruitment, this follows from the increasing recruitment pattern leading to greater numbers of pipeline participants. 
A linear recruitment being one extreme of the mixed recruitment pattern, we observed that, as $l$ increases (i.e., as the recruitment becomes linear for a longer period) a greater EL was observed,with the exact level of EL lying between that under uniform and purely linear $(l = 1)$ recruitment.
Therefore, it can be said that the EL observed assuming linear recruitment may be considered as a maximum possible EL.

We acknowledge that a limitation of our work is that our findings here are based only on a single combination of values for $\alpha$, $\beta$, $\tau$, and $\mu$.
In general, the EL will be dependent on these parameters.
However, the Supplementary Materials contains additional findings that indicate the results altered little when we computed the EL for $\mu = 0.2$. 
In contrast, the EL tended to be lower for $\alpha = 0.01$ (instead of $\alpha = 0.05$), while it inflated further for $\beta = 0.2$ (instead of $\beta = 0.1$).
Finally, the Supplementary Materials indicate how the shape of the stopping boundaries may impact the EL. 
The primary finding was that more aggressive stopping rule translates to larger EL, as it requires a bigger group size at each interim for the same power requirements, thereby impacting the number of pipeline participants.

Note that our work assumes binding futility IAs. 
Non-binding stopping boundaries allow a trial to be continued even after a futility threshold has been crossed, without inflating the type I error rate. 
This increases the complexity of determining the impact of delay on GSDs, as one would need to make assumptions about when the futility rule would be followed/overruled. 
Significantly, the impact of delay may be reduced if it is likely the futility rule will be ignored and the trial subsequently utilise any pipeline participants in later stage analyses. 
That is, binding stopping boundaries can be expected to result in a greater impact of outcome delay, and can be reasonably considered as the worst case scenario at the planning stage of a trial if evaluating the impact of delay.

For unequally spaced IAs, we considered several different possible spacings. 
It was observed that pushing the IAs towards the end of the trial can be beneficial to the expected EG. 
The maximum EL was observed when the first IA was planned even sooner than that under equally spaced IAs. 
When the first IA is pushed toward the end, the EL decreases with respect to a traditional RCT. 
Therefore, the optimal choice for spacing the first IA is largely dependent on the delay length. 
If the delay length is relatively small, a conventional design with equally spaced IAs should work well. 
Whereas, for a large delay length, the EL is reduced if the first IA is conducted towards the latter end of the trial.

We also considered the impact of outcome delay when the optimality criteria is the time to trial completion.
In this case, GSDs always provide more benefit on average vs. a single-stage design, even if it takes a large time to observe the treatment outcome.

In summary, a delay in observing treatment outcomes decreases the expected EG from a GSD in terms of its reduction to the ESS. 
Typically, if the delay length is more than 25\% of the total recruitment period, most of the efficiency gain in terms of ESS is lost due to delay. 
It may be best to use a 2-stage design if the ratio of time to observe the primary outcome to total recruitment period lies below 0.25, as the EL is comparatively lower than for multi-stage GSDs. 
For designs with unequally spaced IAs, pushing the first IA towards the latter end of the trial can be beneficial to the EG.

\section*{Acknowledgements}

JMSW and AM are funded by a NIHR Research Professorship (NIHR301614).

\end{document}


\maketitle
 
\section{Impact of other boundary shapes on efficiency loss}

The following subsection contains results assuming a group-sequential design with Pocock and OBF boundaries. 
The results were obtained assuming $\alpha=0.05, \beta=0.1$ and target treatment effect, $\tau=0.5$.
The total recruitment time was assumed to be 24 months and efficiency lost was plotted for increasing delay length from 1 to 24 months.

\begin{figure}[h]
    \centering
    \includegraphics[width=0.5\textwidth]{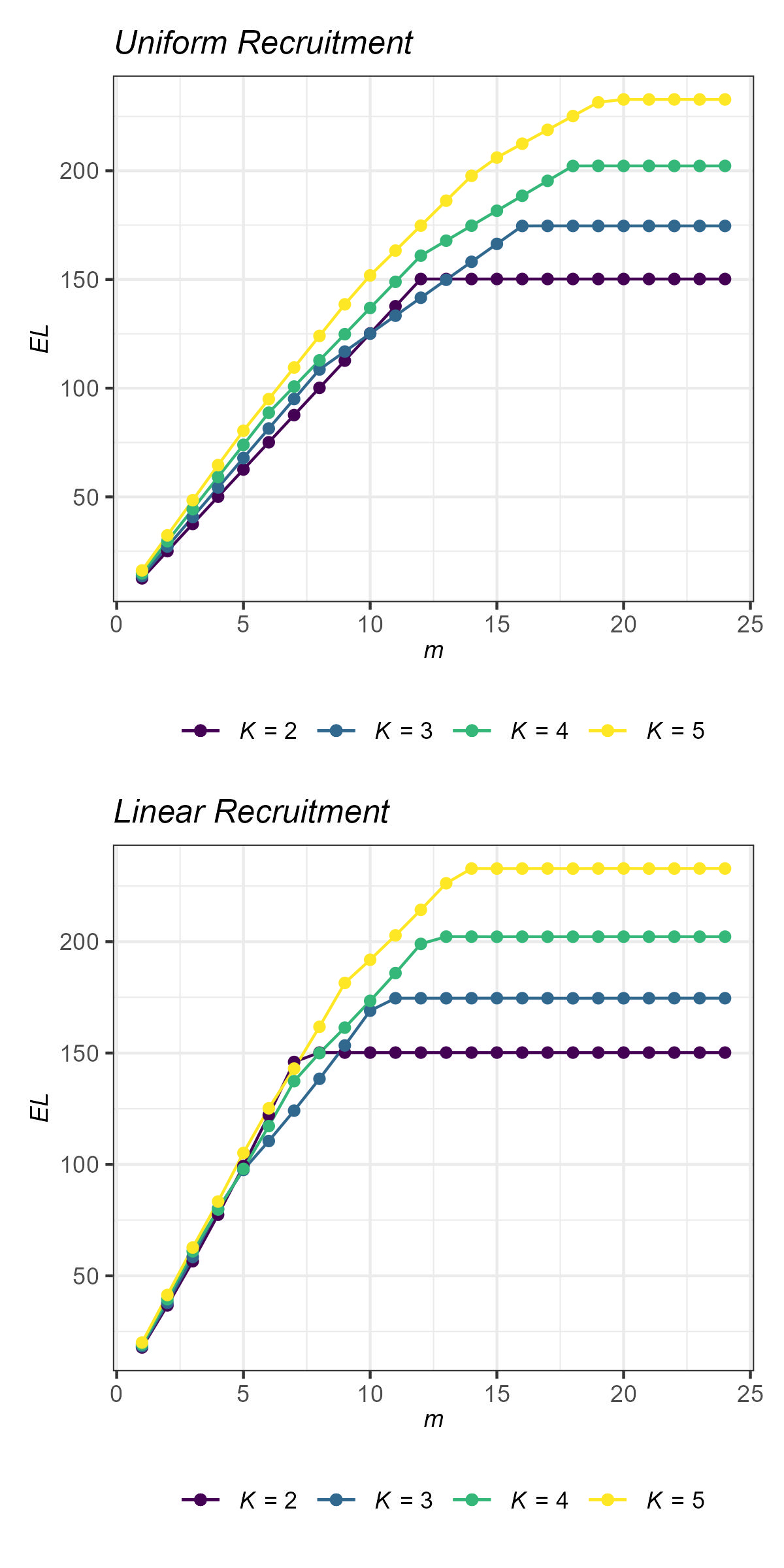}
    \caption{Efficiency lost due to delay for different delay lengths for a GSD with Pocock stopping boundaries assuming uniform and linearly increasing recruitment pattern.\label{SFig1}}
\end{figure}

\begin{figure}[h]
    \centering
    \includegraphics[width=0.5\textwidth]{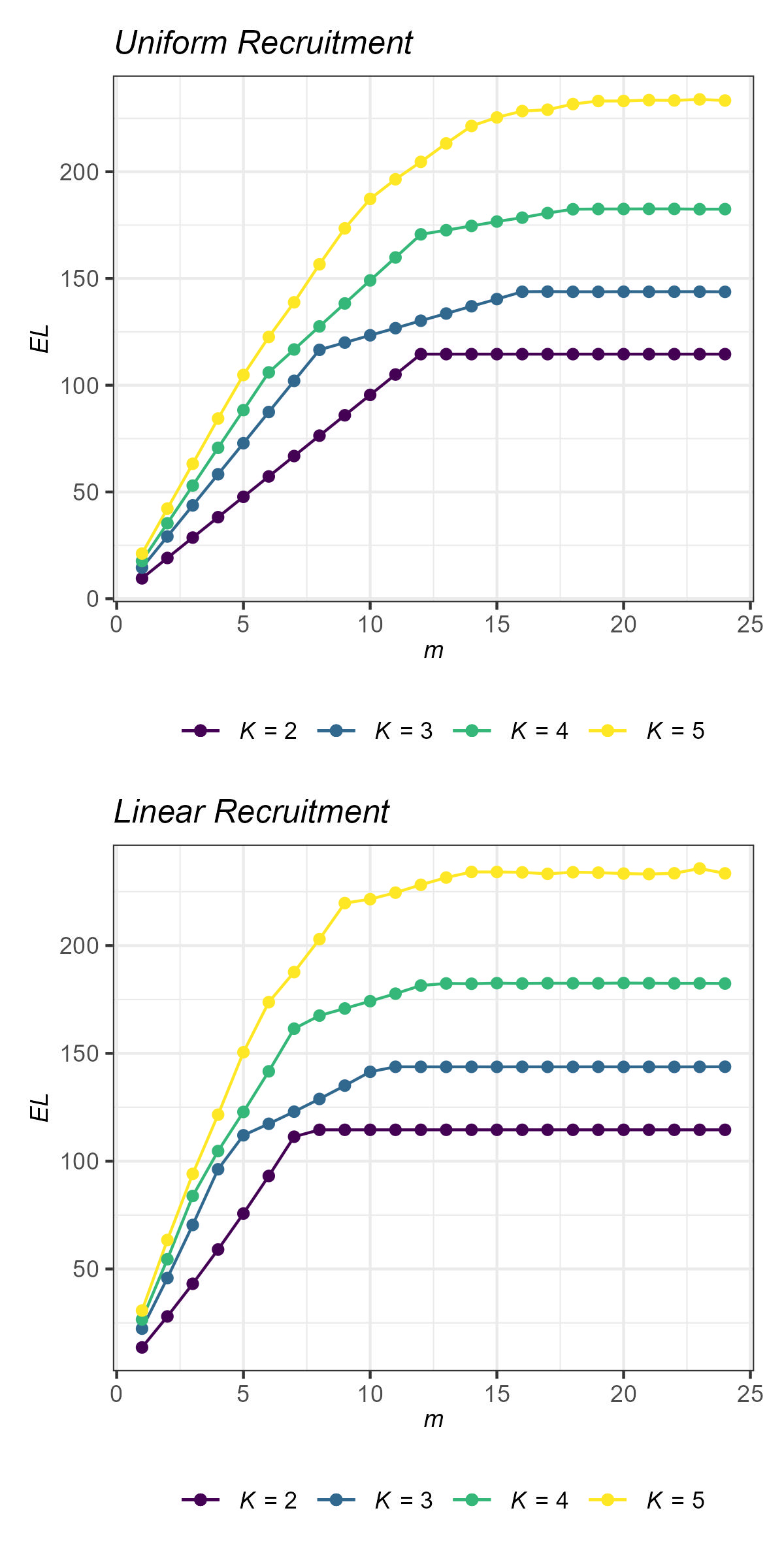}
    \caption{Efficiency lost due to delay for different delay lengths for a GSD with OBF stopping boundaries assuming uniform and linearly increasing recruitment pattern.\label{SFig2}}
\end{figure}

Supplementary Figure~\ref{SFig1} shows the efficiency lost due to delay for Pocock stopping boundaries whereas, Supplementary Figure~\ref{SFig2} shows the same for a O'Brien-Fleming stopping boundary. 
The inferences observed remain similar to the ones obtained using a Wang-Tsiatis boundary. However, we observed that designs with Pocock stopping bounds tend to suffer much greater losses as compared to Wang-Tsiatis or O'Brien-Fleming boundaries.
In general, O'Brien-Fleming designs tend to incur minimum loss compared to the other two stopping boundaries. 

\section{Impact of different type I and type II error values on EL}

This section provides a rough idea on how changing $\alpha$ and $\beta$ values can impact on the efficiency loss values. 

The target treatment effect was assumed to remain the same as in the rest of the paper as, $\tau=0.5$.
The total recruitment time was assumed to be 24 months and efficiency lost was plotted for increasing delay length from 1 to 24 months.

\begin{figure}[h]
    \centering
    \includegraphics[width=0.5\textwidth]{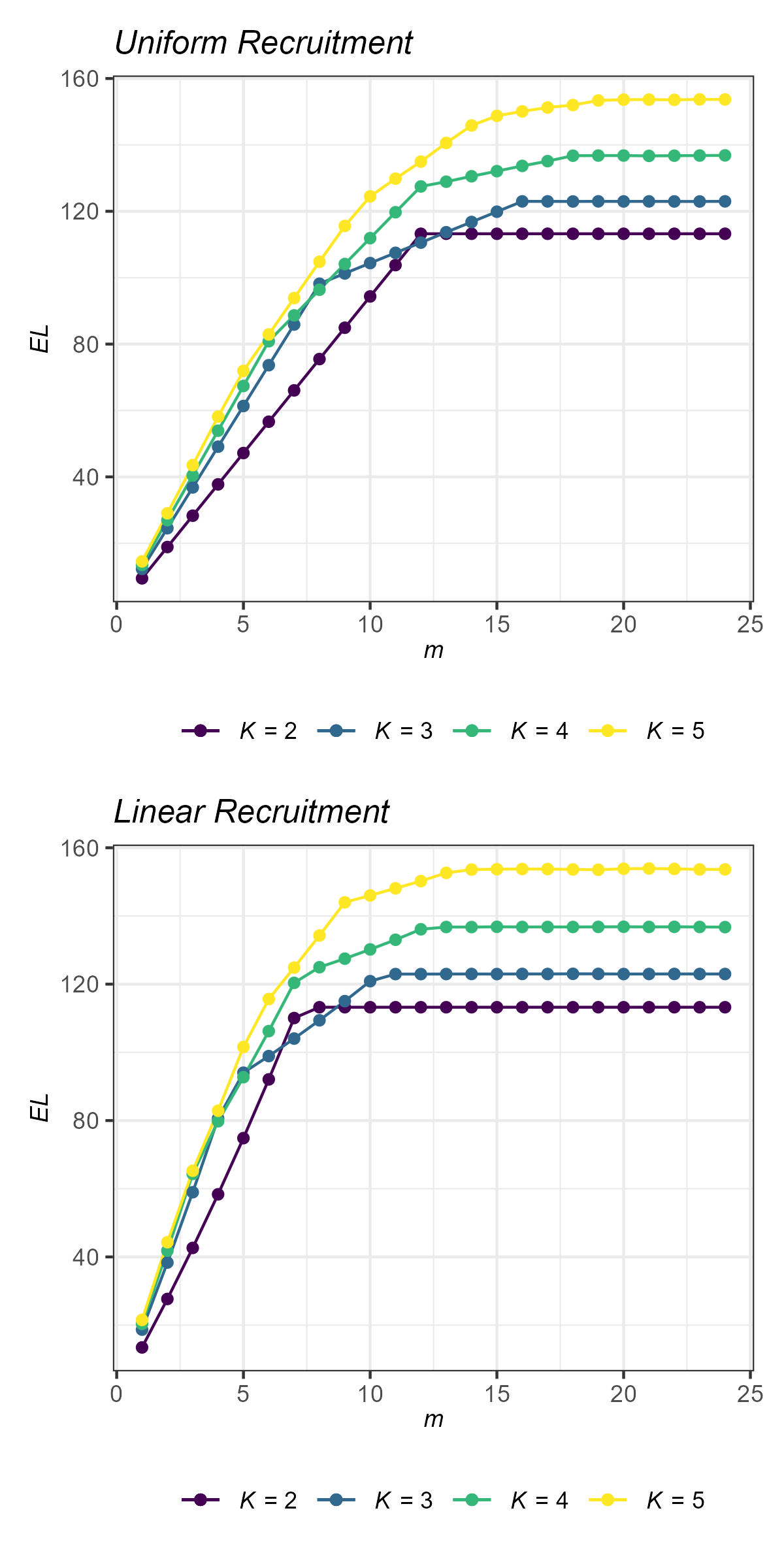}
    \caption{Efficiency lost due to delay assuming $\alpha=0.01$ for different delay lengths for a GSD with WT stopping boundaries assuming uniform and linearly increasing recruitment pattern.\label{Fig alpha}}
\end{figure}

\begin{figure}[h]
    \centering
    \includegraphics[width=0.5\textwidth]{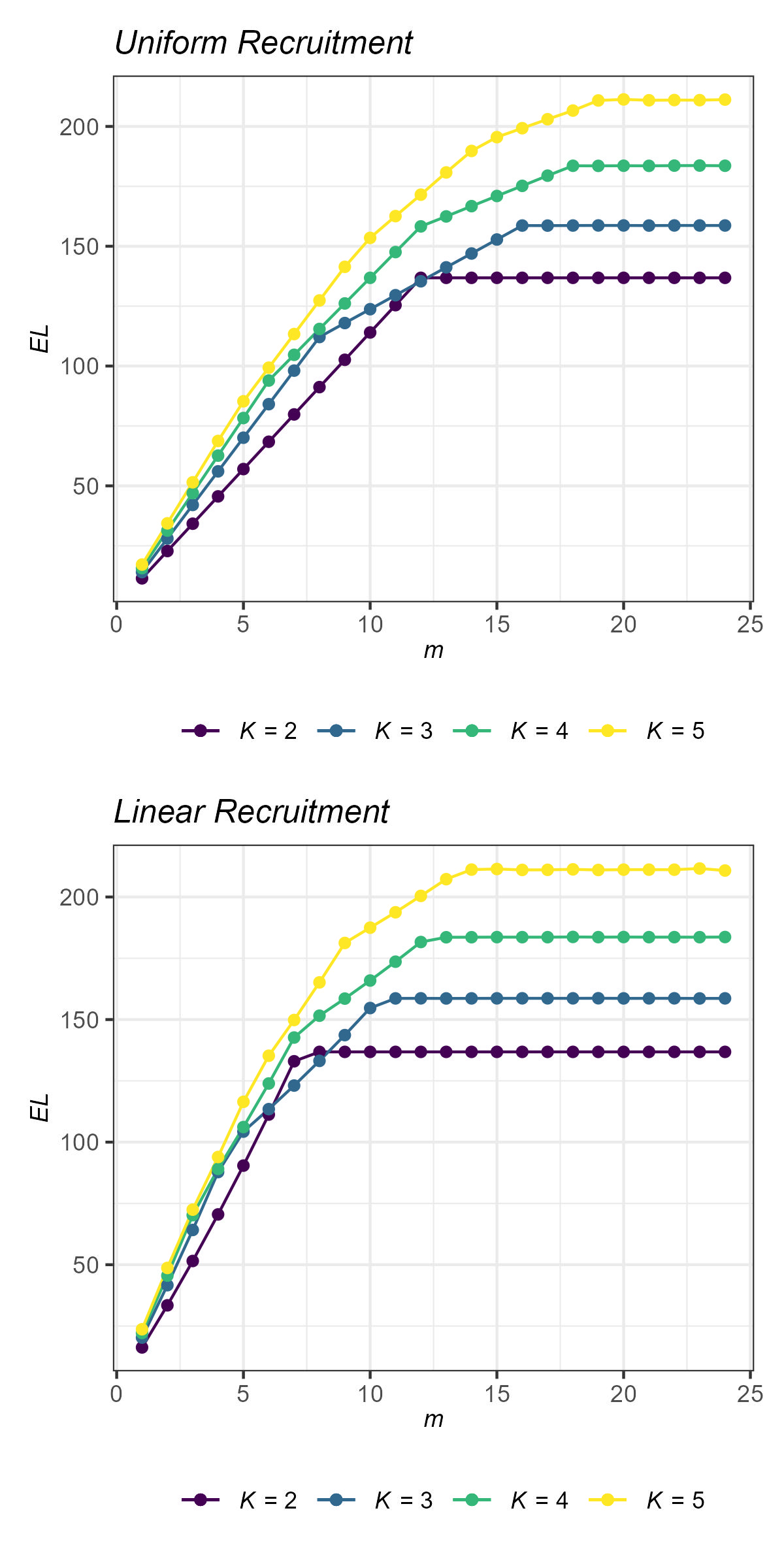}
    \caption{Efficiency lost due to delay assuming $\beta=0.2$ for different delay lengths for a GSD with WT stopping boundaries assuming uniform and linearly increasing recruitment pattern.\label{Fig beta}}
\end{figure}

Supplementary Figure~\ref{Fig alpha} shows the efficiency lost due to delay for $\alpha=0.01$ instead of $\alpha=0.05$.
Supplementary Figure~\ref{Fig beta} shows the efficiency lost for $\beta=0.2$ instead of $\beta=0.1$

\section{Efficiency loss for GSDs with Wang-Tsiatis boundary}

The following tables, Supplementary Table~\ref{Wang-Tsiatis_uni} and Supplementary Table~\ref{Wang-Tsiatis_lin} give a detailed description of the EL values from the result section of the paper. 
The assumptions regarding the parameters of the GSD remains the same as that of the rest of the paper. i.e. we assumed the recruitment pattern to be uniform, linear as well as a mixed recruitment pattern. 
This section also encloses the exact EL values for a GSD with unequally spaced interims for 3 and 4 stage GSD. 

The following list describes the tables given hereafter:

\begin{itemize}
    \item Supplementary Table~\ref{Wang-Tsiatis_uni}: EL assuming a Uniform recruitment pattern.
    \item Supplementary Table~\ref{Wang-Tsiatis_lin}: EL assuming a linear recruitment pattern.
    \item Supplementary Table~\ref{Mixed_2}: EL assuming a mixed recruitment pattern for 2-stage GSD.
    \item Supplementary Table~\ref{Mixed_3}: EL assuming a mixed recruitment pattern for 3-stage GSD.
    \item Supplementary Table~\ref{Mixed_4}: EL assuming a mixed recruitment pattern for 4-stage GSD.
    \item Supplementary Table~\ref{Mixed_5}: EL assuming a mixed recruitment pattern for 5-stage GSD.
    \item Supplementary Table~\ref{Unequal_space_3}: EL assuming an unequally spaced GSD for a 3-stage design.
    \item Supplementary Table~\ref{Unequal_space_4}: EL assuming an unequally spaced GSD for a 4-stage design.
\end{itemize}

\begin{table}[]
\centering
\caption{Efficiency lost under uniform recruitment for a Wang-Tsiatis $(\Delta = 0.25)$ group-sequential design, assuming $\alpha = 0.05,\beta = 0.1$, and $\mu=\tau=0.5$ which give $n_{single}= 137.02$. The total recruitment period is assumed to be 24 months. For each K=2,3,4 and 5, the table records the results for m=3,6,9,12,18 and 24 months respectively.}
\label{Wang-Tsiatis_uni}
\begin{tabular}{cccccccccc}
\bottomrule
$K$ & $n_K$  & $ESS$  & $ESS_{delay}$ &        &        & $\tilde{n}_k$ &       &       & $EL$   \\ \cline{5-9}
    &        &        &               & $k=1$  & $k=2$  & $k=3$         & $k=4$ & $k=5$ &        \\ \bottomrule
2   & 145.05 & 105.84 & 115.64        & 18.13  & 0      &               &       &       & 31.44  \\
    &        &        & 125.45        & 36.26  & 0      &               &       &       & 62.87  \\
    &        &        & 135.25        & 54.39  & 0      &               &       &       & 94.31  \\
    &        &        & 145.05        & 72.52  & 0      &               &       &       & 125.75 \\
    &        &        & 145.05        & 72.52  & 0      &               &       &       & 125.75 \\
    &        &        & 145.05        & 72.52  & 0      &               &       &       & 125.75 \\ \hline
3   & 155.57 & 98.74  & 113.60        & 19.45  & 19.45  & 0             &       &       & 38.82  \\
    &        &        & 128.47        & 38.89  & 38.89  & 0             &       &       & 77.66  \\
    &        &        & 140.52        & 58.34  & 51.86  & 0             &       &       & 109.14 \\
    &        &        & 146.97        & 77.79  & 51.86  & 0             &       &       & 125.99 \\
    &        &        & 155.58        & 103.72 & 51.86  & 0             &       &       & 148.48 \\
    &        &        & 155.58        & 103.72 & 51.86  & 0             &       &       & 148.47 \\ \hline
4   & 169.14 & 95.98  & 114.27        & 21.14  & 21.14  & 21.14         & 0     &       & 44.57  \\
    &        &        & 132.56        & 42.28  & 42.28  & 42.28         & 0     &       & 89.12  \\
    &        &        & 145.90        & 63.43  & 63.43  & 42.28         & 0     &       & 121.64 \\
    &        &        & 159.25        & 84.57  & 84.57  & 42.28         & 0     &       & 154.15 \\
    &        &        & 169.15        & 126.86 & 84.58  & 42.29         & 0     &       & 178.30 \\
    &        &        & 169.14        & 126.85 & 84.57  & 42.28         & 0     &       & 178.24 \\ \hline
5   & 185.23 & 95.09  & 116.39        & 23.15  & 23.15  & 23.15         & 23.15 & 0     & 50.80  \\
    &        &        & 136.54        & 46.31  & 46.31  & 46.31         & 37.05 & 0     & 98.84  \\
    &        &        & 154.90        & 69.46  & 69.46  & 69.46         & 37.04 & 0     & 142.63 \\
    &        &        & 168.45        & 92.60  & 92.60  & 74.08         & 37.04 & 0     & 174.94 \\
    &        &        & 183.52        & 138.93 & 111.14 & 74.10         & 37.05 & 0     & 210.92 \\
    &        &        & 185.20        & 148.16 & 111.12 & 74.08         & 37.04 & 0     & 214.87\\
    \bottomrule
\end{tabular}
\end{table}

\begin{table}[]
\centering
\caption{Efficiency lost under linear recruitment for a Wang-Tsiatis $(\Delta = 0.25)$ group-sequential design, assuming $\alpha = 0.05,\beta = 0.1$, and $\mu=\tau=0.5$ which give $n_{single}= 137.02$. The total recruitment period is assumed to be 24 months. For each K=2,3,4 and 5, the table records the results for m=3,6,9,12,18 and 24 months respectively.}
\label{Wang-Tsiatis_lin}
\begin{tabular}{cccccccccc}
\bottomrule
$K$ & $n_K$  & $ESS$  & $ESS_{delay}$ &        &        & $\tilde{n}_k$ &       &       & $EL$   \\ \cline{5-9}
    &        &        &               & $k=1$  & $k=2$  & $k=3$         & $k=4$ & $k=5$ &        \\ \bottomrule
2 & 145.05 & 105.84 & 120.61 & 27.31  & 0      &       &       &   & 47.35  \\
  &        &        & 137.72 & 58.97  & 0      &       &       &   & 102.25 \\
  &        &        & 145.05 & 72.52  & 0      &       &       &   & 125.75 \\
  &        &        & 145.05 & 72.52  & 0      &       &       &   & 125.75 \\
  &        &        & 145.05 & 72.52  & 0      &       &       &   & 125.75 \\
  &        &        & 145.05 & 72.52  & 0      &       &       &   & 125.75 \\ \hline
3 & 155.58 & 98.74  & 121.29 & 24.35  & 33.46  & 0     &       &   & 58.91  \\
  &        &        & 138.87 & 53.36  & 51.86  & 0     &       &   & 104.83 \\
  &        &        & 150.04 & 87.05  & 51.86  & 0     &       &   & 134.01 \\
  &        &        & 155.57 & 103.71 & 51.86  & 0     &       &   & 148.44 \\
  &        &        & 155.57 & 103.72 & 51.86  & 0     &       &   & 148.46 \\
  &        &        & 155.58 & 103.72 & 51.86  & 0     &       &   & 148.47 \\ \hline
4 & 169.12 & 95.97  & 123.05 & 23.27  & 31.84  & 38.42 & 0     &   & 65.97  \\
  &        &        & 145.27 & 51.61  & 68.77  & 42.29 & 0     &   & 120.10 \\
  &        &        & 159.38 & 85.04  & 84.58  & 42.29 & 0     &   & 154.48 \\
  &        &        & 168.36 & 123.52 & 84.57  & 42.28 & 0     &   & 176.34 \\
  &        &        & 169.13 & 126.85 & 84.56  & 42.28 & 0     &   & 178.22 \\
  &        &        & 169.14 & 126.85 & 84.57  & 42.28 & 0     &   & 178.24 \\ \hline
5 & 185.24 & 95.10  & 124.89 & 23.09  & 31.49  & 37.94 & 37.05 & 0 & 71.06  \\
  &        &        & 152.52 & 51.74  & 68.54  & 74.10 & 37.05 & 0 & 136.97 \\
  &        &        & 173.70 & 85.94  & 111.14 & 74.09 & 37.05 & 0 & 187.49 \\
  &        &        & 181.07 & 125.70 & 111.14 & 74.09 & 37.05 & 0 & 205.06 \\
  &        &        & 185.26 & 148.21 & 111.16 & 74.10 & 37.05 & 0 & 215.10 \\
  &        &        & 185.22 & 148.18 & 111.13 & 74.09 & 37.04 & 0 & 214.95\\\bottomrule
\end{tabular}
\end{table}

\begin{table}[htbp]
\centering
\caption{Efficiency lost under a mixed recruitment pattern for a 2-stage Wang-Tsiatis ($\Delta = 0.25$) group-sequential design, assuming $\alpha = 0.025$, $\beta = 0.1$, and $\mu=\tau=0.5$ which give $n_\text{single} = 168.12$.
The total recruitment period is assumed to be 24 months.
The table records the results for $m=3,6,9,12,18$ and $24$ months respectively.
The maximum sample size for the group-sequential design is 173.86.
Here, $l$ takes values 0.2, 0.4, 0.6 and 0.8 to denote the increasing degree of linearity in the recruitment pattern.}
\label{Mixed_2}
\begin{tabular}{ccccccc@{}}
\toprule
 $m$& $l$& $ESS$& $ESS_{delay}$&$\tilde{n}_1$ &$\tilde{n}_2$ &$EL$ \\\bottomrule
 3  & 0.2 & 105.84 & 116.30 & 19.34 & 0 & 33.53  \\
   & 0.4 &        & 117.61 & 21.76 & 0 & 37.72  \\
   & 0.6 &        & 119.29 & 24.87 & 0 & 43.11  \\
   & 0.8 &        & 121.53 & 29.01 & 0 & 50.30  \\
   \hline
6  & 0.2 &        & 126.75 & 38.68 & 0 & 67.06  \\
   & 0.4 &        & 129.37 & 43.51 & 0 & 75.45  \\
   & 0.6 &        & 132.73 & 49.73 & 0 & 86.23  \\
   & 0.8 &        & 137.21 & 58.02 & 0 & 100.60 \\
   \hline
9  & 0.2 &        & 137.21 & 58.02 & 0 & 100.60 \\
   & 0.4 &        & 141.13 & 65.27 & 0 & 113.17 \\
   & 0.6 &        & 145.05 & 72.52 & 0 & 125.75 \\
   & 0.8 &        & 145.05 & 72.52 & 0 & 125.75 \\
   \hline
12 & 0.2 &        & 145.05 & 72.52 & 0 & 125.75 \\
   & 0.4 &        & 145.05 & 72.52 & 0 & 125.75 \\
   & 0.6 &        & 145.05 & 72.52 & 0 & 125.75 \\
   & 0.8 &        & 145.05 & 72.52 & 0 & 125.75 \\
   \hline
18 & 0.2 &        & 145.05 & 72.52 & 0 & 125.75 \\
   & 0.4 &        & 145.05 & 72.52 & 0 & 125.75 \\
   & 0.6 &        & 145.05 & 72.52 & 0 & 125.75 \\
   & 0.8 &        & 145.05 & 72.52 & 0 & 125.75 \\
   \hline
24 & 0.2 &        & 145.05 & 72.52 & 0 & 125.75 \\
   & 0.4 &        & 145.05 & 72.52 & 0 & 125.75 \\
   & 0.6 &        & 145.05 & 72.52 & 0 & 125.75 \\
   & 0.8 &        & 145.05 & 72.52 & 0 & 125.75 \\ \bottomrule
\end{tabular}
\end{table}

\begin{table}[htbp]
\centering
\caption{Efficiency lost under a mixed recruitment pattern for a 3-stage Wang-Tsiatis ($\Delta = 0.25$) group-sequential design, assuming $\alpha = 0.025$, $\beta = 0.1$, and $\mu=\tau=0.5$ which give $n_\text{single} = 168.12$.
The total recruitment period is assumed to be 24 months.
The table records the results for $m=3,6,9,12,18$ and $24$ months respectively.
The maximum sample size for the group-sequential design is 176.49
Here, $l$ takes values 0.2, 0.4, 0.6 and 0.8 to denote the increasing degree of linearity in the recruitment pattern.}
\label{Mixed_3}
\begin{tabular}{cccccccc@{}}
\toprule
$m$& $l$& $ESS$& $ESS_{delay}$&$\tilde{n}_1$ &$\tilde{n}_2$ &$\tilde{n}_3$&$EL$ \\\bottomrule
 3  & 0.2 & 98.73 & 114.59 & 20.74  & 20.74 & 0 & 41.40  \\
   & 0.4 &       & 116.58 & 23.34  & 23.34 & 0 & 46.59  \\
   & 0.6 &       & 118.14 & 23.71  & 26.67 & 0 & 50.67  \\
   & 0.8 &       & 120.35 & 24.57  & 31.12 & 0 & 56.45  \\
   \hline
6  & 0.2 &       & 130.45 & 41.49  & 41.49 & 0 & 82.83  \\
   & 0.4 &       & 134.41 & 46.67  & 46.67 & 0 & 93.17  \\
   & 0.6 &       & 137.89 & 50.38  & 51.86 & 0 & 102.26 \\
   & 0.8 &       & 138.37 & 51.86  & 51.86 & 0 & 103.53 \\
   \hline
9  & 0.2 &       & 141.81 & 62.23  & 51.86 & 0 & 112.51 \\
   & 0.4 &       & 144.39 & 70.01  & 51.86 & 0 & 119.24 \\
   & 0.6 &       & 146.73 & 77.05  & 51.86 & 0 & 125.36 \\
   & 0.8 &       & 148.69 & 82.97  & 51.86 & 0 & 130.49 \\
   \hline
12 & 0.2 &       & 148.69 & 82.97  & 51.86 & 0 & 130.47 \\
   & 0.4 &       & 152.14 & 93.35  & 51.86 & 0 & 139.48 \\
   & 0.6 &       & 155.58 & 103.72 & 51.86 & 0 & 148.47 \\
   & 0.8 &       & 155.58 & 103.72 & 51.86 & 0 & 148.48 \\
   \hline
18 & 0.2 &       & 155.58 & 103.72 & 51.86 & 0 & 148.47 \\
   & 0.4 &       & 155.58 & 103.72 & 51.86 & 0 & 148.47 \\
   & 0.6 &       & 155.57 & 103.71 & 51.86 & 0 & 148.45 \\
   & 0.8 &       & 155.58 & 103.72 & 51.86 & 0 & 148.47 \\
   \hline
24 & 0.2 &       & 155.57 & 103.72 & 51.86 & 0 & 148.46 \\
   & 0.4 &       & 155.58 & 103.72 & 51.86 & 0 & 148.48 \\
   & 0.6 &       & 155.58 & 103.72 & 51.86 & 0 & 148.48 \\
   & 0.8 &       & 155.57 & 103.71 & 51.86 & 0 & 148.46 \\ \bottomrule
\end{tabular}
\end{table}

\begin{table}[htbp]
\centering
\caption{Efficiency lost under a mixed recruitment pattern for a 4-stage Wang-Tsiatis ($\Delta = 0.25$) group-sequential design, assuming $\alpha = 0.025$, $\beta = 0.1$, and $\mu=\tau=0.5$ which give $n_\text{single} = 168.12$.
The total recruitment period is assumed to be 24 months.
The table records the results for $m=3,6,9,12,18$ and $24$ months respectively.
The maximum sample size for the group-sequential design is 178.12
Here, $l$ takes values 0.2, 0.4, 0.6 and 0.8 to denote the increasing degree of linearity in the recruitment pattern.}
\label{Mixed_4}
\begin{tabular}{ccccccccc@{}}
\toprule
$m$& $l$& $ESS$& $ESS_{delay}$&$\tilde{n}_1$ &$\tilde{n}_2$ &$\tilde{n}_3$ &$\tilde{n}_4$& $EL$ \\\bottomrule
3  & 0.2 & 98.73 & 114.59 & 20.74  & 20.74 & 0 & 41.40  \\
   & 0.4 &       & 116.58 & 23.34  & 23.34 & 0 & 46.59  \\
   & 0.6 &       & 118.14 & 23.71  & 26.67 & 0 & 50.67  \\
   & 0.8 &       & 120.35 & 24.57  & 31.12 & 0 & 56.45  \\ \hline
6  & 0.2 &       & 130.45 & 41.49  & 41.49 & 0 & 82.83  \\
   & 0.4 &       & 134.41 & 46.67  & 46.67 & 0 & 93.17  \\
   & 0.6 &       & 137.89 & 50.38  & 51.86 & 0 & 102.26 \\
   & 0.8 &       & 138.37 & 51.86  & 51.86 & 0 & 103.53 \\ \hline
9  & 0.2 &       & 141.81 & 62.23  & 51.86 & 0 & 112.51 \\
   & 0.4 &       & 144.39 & 70.01  & 51.86 & 0 & 119.24 \\
   & 0.6 &       & 146.73 & 77.05  & 51.86 & 0 & 125.36 \\
   & 0.8 &       & 148.69 & 82.97  & 51.86 & 0 & 130.49 \\ \hline
12 & 0.2 &       & 148.69 & 82.97  & 51.86 & 0 & 130.47 \\
   & 0.4 &       & 152.14 & 93.35  & 51.86 & 0 & 139.48 \\
   & 0.6 &       & 155.58 & 103.72 & 51.86 & 0 & 148.47 \\
   & 0.8 &       & 155.58 & 103.72 & 51.86 & 0 & 148.48 \\ \hline
18 & 0.2 &       & 155.58 & 103.72 & 51.86 & 0 & 148.47 \\
   & 0.4 &       & 155.58 & 103.72 & 51.86 & 0 & 148.47 \\
   & 0.6 &       & 155.57 & 103.71 & 51.86 & 0 & 148.45 \\
   & 0.8 &       & 155.58 & 103.72 & 51.86 & 0 & 148.47 \\ \hline
24 & 0.2 &       & 155.57 & 103.72 & 51.86 & 0 & 148.46 \\
   & 0.4 &       & 155.58 & 103.72 & 51.86 & 0 & 148.48 \\
   & 0.6 &       & 155.58 & 103.72 & 51.86 & 0 & 148.48 \\
   & 0.8 &       & 155.57 & 103.71 & 51.86 & 0 & 148.46\\\bottomrule
\end{tabular}
\end{table}

\begin{table}[htbp]
\centering
\caption{Efficiency lost under a mixed recruitment pattern for a 5-stage Wang-Tsiatis ($\Delta = 0.25$) group-sequential design, assuming $\alpha = 0.025$, $\beta = 0.1$, and $\mu=\tau=0.5$ which give $n_\text{single} = 168.12$.
The total recruitment period is assumed to be 24 months.
The table records the results for $m=3,6,9,12,18$ and $24$ months respectively.
The maximum sample size for the group-sequential design is 179.25.
Here, $l$ takes values 0.2, 0.4, 0.6 and 0.8 to denote the increasing degree of linearity in the recruitment pattern.}
\label{Mixed_5}
\begin{tabular}{cccccccccc@{}}
\toprule
$m$& $l$& $ESS$& $ESS_{delay}$&$\tilde{n}_1$ &$\tilde{n}_2$ &$\tilde{n}_3$&$\tilde{n}_4$& $\tilde{n}_5$&$EL$ \\\bottomrule
 3  & 0.2 & 95.10 & 117.82 & 24.70  & 24.70  & 24.70 & 24.70 & 0 & 54.20  \\
   & 0.4 &       & 120.64 & 27.78  & 27.78  & 27.78 & 27.78 & 0 & 60.94  \\
   & 0.6 &       & 124.89 & 24.95  & 37.05  & 31.76 & 31.76 & 0 & 71.08  \\
   & 0.8 &       & 125.30 & 23.40  & 33.15  & 37.05 & 37.05 & 0 & 72.03  \\
   \hline
6  & 0.2 &       & 139.00 & 49.40  & 49.40  & 49.40 & 37.05 & 0 & 104.72 \\
   & 0.4 &       & 143.86 & 55.56  & 55.56  & 55.56 & 37.04 & 0 & 116.32 \\
   & 0.6 &       & 150.77 & 56.71  & 68.80  & 63.51 & 37.05 & 0 & 132.79 \\
   & 0.8 &       & 153.92 & 52.64  & 72.14  & 74.09 & 37.04 & 0 & 140.29 \\
   \hline
9  & 0.2 &       & 158.58 & 74.09  & 74.09  & 74.09 & 37.05 & 0 & 151.43 \\
   & 0.4 &       & 163.54 & 83.36  & 83.36  & 74.10 & 37.05 & 0 & 163.27 \\
   & 0.6 &       & 171.74 & 95.26  & 100.56 & 74.09 & 37.05 & 0 & 182.82 \\
   & 0.8 &       & 174.03 & 87.74  & 111.14 & 74.09 & 37.05 & 0 & 188.28 \\
   \hline
12 & 0.2 &       & 171.78 & 98.79  & 98.79  & 74.09 & 37.05 & 0 & 182.90 \\
   & 0.4 &       & 178.38 & 111.15 & 111.15 & 74.10 & 37.05 & 0 & 198.67 \\
   & 0.6 &       & 183.84 & 140.63 & 111.14 & 74.10 & 37.05 & 0 & 211.68 \\
   & 0.8 &       & 181.63 & 128.70 & 111.15 & 74.10 & 37.05 & 0 & 206.42 \\
   \hline
18 & 0.2 &       & 185.21 & 148.17 & 111.13 & 74.08 & 37.04 & 0 & 214.92 \\
   & 0.4 &       & 185.23 & 148.19 & 111.14 & 74.09 & 37.05 & 0 & 214.99 \\
   & 0.6 &       & 185.21 & 148.16 & 111.12 & 74.08 & 37.04 & 0 & 214.91 \\
   & 0.8 &       & 185.23 & 148.18 & 111.14 & 74.09 & 37.05 & 0 & 214.97 \\
   \hline
24 & 0.2 &       & 185.23 & 148.18 & 111.14 & 74.09 & 37.05 & 0 & 214.99 \\
   & 0.4 &       & 185.22 & 148.18 & 111.13 & 74.09 & 37.04 & 0 & 214.94 \\
   & 0.6 &       & 185.22 & 148.18 & 111.13 & 74.09 & 37.04 & 0 & 214.96 \\
   & 0.8 &       & 185.23 & 148.18 & 111.14 & 74.09 & 37.05 & 0 & 214.97\\ \bottomrule
\end{tabular}
\end{table}

\begin{table}[htbp]
\centering
\caption{Efficiency lost for a unequally spaced group-sequential design with $K=3$ for uniform recruitment.
The designs recorded assumes $\alpha = 0.025$, $\beta = 0.1$, and $\mu=\tau=0.5$ which also gives the equivalent $n_\text{single} = 168.12$.
The total recruitment period is assumed to be 24 months.
The table records the results for $m=3,6,9,12,18$ and $24$ months respectively.
Here I represents equally spaced interims;
II represents the first interims takes place sooner, $(0.25, 0.5, 1)$;
III represents the first interims take place later, $(0.5, 0.75, 1)$; and
IV represents the first interims occur even later, after 60\% and 90\% of the total recruitment, $(0.6, 0.9, 1)$.}
\label{Unequal_space_3}
\begin{tabular}{@{}ccccccccc@{}}
\toprule
$m$& \textit{Interim Spacing}& $n_K$ &$ESS$& $ESS_{delay}$&$\tilde{n}_1$ &$\tilde{n}_2$ &$\tilde{n}_3$&$EL$ \\\bottomrule
3  & I   & 155.58 & 98.74  & 113.60 & 19.45  & 19.45 & 0 & 38.83  \\
   & II  & 167.56 & 101.77 & 115.54 & 20.95  & 20.95 & 0 & 39.04  \\
   & III & 147.63 & 99.60  & 114.11 & 18.45  & 18.45 & 0 & 38.77  \\
   & IV  & 144.80 & 105.48 & 120.20 & 18.10  & 14.48 & 0 & 46.67  \\ \hline
6  & I   & 155.58 & 98.74  & 128.47 & 38.89  & 38.89 & 0 & 77.65  \\
   & II  & 167.59 & 101.79 & 129.32 & 41.90  & 41.90 & 0 & 78.14  \\
   & III & 147.64 & 99.61  & 128.63 & 36.91  & 36.91 & 0 & 77.57  \\
   & IV  & 144.80 & 105.48 & 131.38 & 36.20  & 14.48 & 0 & 82.12  \\ \hline
9  & I   & 155.58 & 98.74  & 140.53 & 58.34  & 51.86 & 0 & 109.15 \\
   & II  & 167.57 & 101.78 & 143.07 & 62.84  & 62.84 & 0 & 117.15 \\
   & III & 147.63 & 99.61  & 138.13 & 55.36  & 36.91 & 0 & 102.96 \\
   & IV  & 144.80 & 105.48 & 142.56 & 54.30  & 14.48 & 0 & 117.57 \\ \hline
12 & I   & 155.57 & 98.74  & 146.97 & 77.79  & 51.86 & 0 & 125.99 \\
   & II  & 167.58 & 101.79 & 156.84 & 83.79  & 83.79 & 0 & 156.25 \\
   & III & 147.63 & 99.61  & 147.63 & 73.82  & 36.91 & 0 & 128.37 \\
   & IV  & 144.80 & 105.48 & 144.80 & 57.92  & 14.48 & 0 & 124.66 \\ \hline
18 & I   & 155.58 & 98.74  & 155.58 & 103.72 & 51.86 & 0 & 148.47 \\
   & II  & 167.59 & 101.80 & 167.59 & 125.69 & 83.79 & 0 & 186.78 \\
   & III & 147.64 & 99.61  & 147.64 & 73.82  & 36.91 & 0 & 128.37 \\
   & IV  & 144.80 & 105.48 & 144.80 & 57.92  & 14.48 & 0 & 124.67 \\ \hline
24 & I   & 155.57 & 98.74  & 155.57 & 103.72 & 51.86 & 0 & 148.47 \\
   & II  & 167.57 & 101.78 & 167.57 & 125.68 & 83.79 & 0 & 186.69 \\
   & III & 147.64 & 99.61  & 147.64 & 73.82  & 36.91 & 0 & 128.38 \\
   & IV  & 144.80 & 105.48 & 144.80 & 57.92  & 14.48 & 0 & 124.66\\ \bottomrule
\end{tabular}
\end{table}

\begin{table}[htbp]
\centering
\caption{Efficiency lost for a unequally spaced group-sequential design with $K=4$ for uniform recruitment.
The designs recorded assumes $\alpha = 0.025$, $\beta = 0.1$, and $\mu=\tau=0.5$ which also gives the equivalent $n_\text{single} = 168.12$.
The total recruitment period is assumed to be 24 months.
The table records the results for $m=3,6,9,12,18$ and $24$ months respectively.
Here I represents equally spaced interims; II represents interims done at 20, 40, 60 and 100\% of the total sample size, i.e. the first interim is done sooner than an equally spaced design; III represents interims done at 40,60,80 and 100\% of the total sample size, i.e. the first and subsequent interims are pushed to the latter end of the design.}
\label{Unequal_space_4}
\begin{tabular}{@{}cccccccccc@{}}
\toprule
$m$& \textit{Interim Spacing}& $n_K$ &$ESS$& $ESS_{delay}$&$\tilde{n}_1$ &$\tilde{n}_2$ &$\tilde{n}_3$ &$\tilde{n}_4$ &$EL$ \\\bottomrule
3  & I   & 169.16 & 95.99 & 114.28 & 21.14  & 21.14  & 21.14 & 0 & 44.58  \\
   & II  & 184.61 & 96.99 & 115.65 & 23.08  & 23.08  & 23.08 & 0 & 46.62  \\
   & III & 152.61 & 94.77 & 110.80 & 19.08  & 19.08  & 19.08 & 0 & 37.93  \\ \hline
6  & I   & 169.14 & 95.98 & 132.56 & 42.29  & 42.29  & 42.29 & 0 & 89.13  \\
   & II  & 184.61 & 96.98 & 134.31 & 46.15  & 46.15  & 46.15 & 0 & 93.23  \\
   & III & 152.60 & 94.77 & 125.50 & 38.15  & 38.15  & 30.52 & 0 & 72.73  \\ \hline
9  & I   & 169.12 & 95.97 & 145.89 & 63.42  & 63.42  & 42.28 & 0 & 121.60 \\
   & II  & 184.59 & 96.97 & 152.96 & 69.22  & 69.22  & 69.22 & 0 & 139.79 \\
   & III & 152.61 & 94.77 & 138.23 & 57.23  & 57.23  & 30.52 & 0 & 102.85 \\ \hline
12 & I   & 169.14 & 95.98 & 159.25 & 84.57  & 84.57  & 42.28 & 0 & 154.17 \\
   & II  & 184.59 & 96.98 & 166.96 & 92.30  & 92.30  & 73.84 & 0 & 174.77 \\
   & III & 152.62 & 94.78 & 146.70 & 76.31  & 61.05  & 30.52 & 0 & 122.91 \\ \hline
18 & I   & 169.14 & 95.98 & 169.14 & 126.86 & 84.57  & 42.29 & 0 & 178.26 \\
   & II  & 184.58 & 96.97 & 182.74 & 138.43 & 110.75 & 73.83 & 0 & 214.14 \\
   & III & 152.59 & 94.76 & 152.59 & 91.55  & 61.04  & 30.52 & 0 & 136.84 \\ \hline
24 & I   & 169.14 & 95.97 & 169.14 & 126.85 & 84.57  & 42.28 & 0 & 178.24 \\
   & II  & 184.59 & 96.97 & 184.59 & 147.67 & 110.75 & 73.83 & 0 & 218.77 \\
   & III & 152.61 & 94.77 & 152.61 & 91.56  & 61.04  & 30.52 & 0 & 136.89\\\bottomrule
\end{tabular}
\end{table}

\section{Results assuming symmetric stopping boundaries for equally spaced IAs}

\begin{figure}
     \centering
     \includegraphics[width=0.5\linewidth]{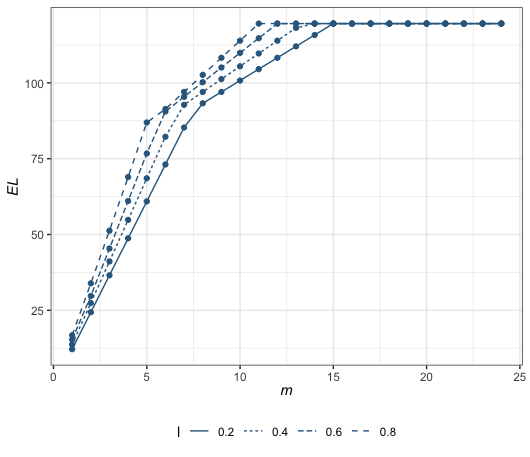}
     \caption{Efficiency loss (EL) due to delay, for different delay lengths $m$, assuming equally spaced interim analyses in a 3-stage design, under a mixed recruitment pattern categorised by specific values of $l$.}
     \label{sup.fig5}
 \end{figure}

The following section provides results for the EL assuming a symmetric stopping boundary. Here, the efficacy and futility boundaries were chosen assuming the Wang-Tsiatis boundary with shape parameter 0.25. 
The following plot shows the EL varying across delay length for a mixed recruitment pattern.
For Supplementary Figure~\ref{sup.fig5}, it can be observed that the results obtained align with our previous findings, i.e., as the recruitment pattern becomes linear for a greater proportion of the total recruitment time, the EL increases.

\section{Results assuming symmetric stopping boundaries for unequally spaced IAs}

Here, the results are obtained with interim spacings as mention in section 3.2 in the main paper. The stopping boundaries used are obtained using a Hwang Shih decani spending functions with parameter -2.

\begin{figure}
    \centering
    \includegraphics[width=0.5\linewidth]{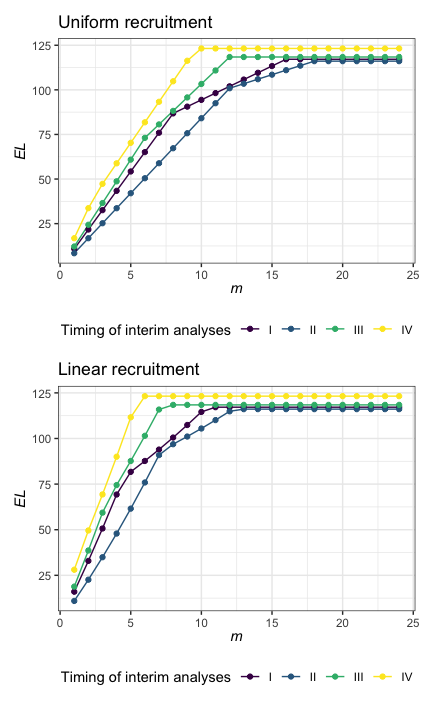}
    \caption{Efficiency loss (EL) due to delay for different delay lengths, in 3-stage designs with unequally spaced interim analyses, under uniform and linear recruitment patterns.}
    \label{sup.fig6}
\end{figure}

It can be observed from Supplementary Figure \ref{sup.fig6} that, in general, a GSD with equally spaced IAs performs better than a GSD with unequally spaced IAs. 
The exception to this is the case when the first IA is performed even sooner than that under equal spacing, i.e., the design where the two IAs are conducted at 25\% and 50\% of the total sample size incurs the smallest losses. 
If the first and subsequent IAs are pushed to the latter end of the trial the efficiency loss increases rapidly with the delay length. 
This is because, as we push the interims towards the latter end of the trial, the maximum sample size increases, thereby increasing the recruitment rate based on the assumptions of the recruitment model. 
For a linearly increasing recruitment this is further influenced by the factor that towards the end of the trial, there is a larger numbers of pipeline participants (as the recruitment pattern is increasing in nature). 
This inflates $ESS_{delay}$ and thus also the EL.
We observe that the EL crosses 100\% for IA timings (0.6, 0.9, 1) for $m = 8$ in contrast to $m = 12$ for equally spaced interims. 
This EL occurs even sooner at a 5 month delay, rather than 8 months, under linear recruitment. 
A maximal EL of 123.20\% is observed for the IA timings (0.6, 0.9, 1) when the delay length is more than 10 months; this EL occurs even sooner at 6 months delay, under linear recruitment.